\documentclass[conference]{IEEEtran}
\IEEEoverridecommandlockouts
\usepackage[utf8]{inputenc}
\usepackage[T1]{fontenc}
\usepackage{graphicx}
\usepackage{subcaption}
\usepackage{cite}
\usepackage{amsmath,amssymb,amsfonts}
\usepackage{algorithmic}
\usepackage{graphicx}
\usepackage{textcomp}
\usepackage{xcolor}
\usepackage{float}

\usepackage{xcolor,soul}
\usepackage{xspace}
\usepackage{enumitem}
\usepackage{titlecaps}
\usepackage{tabularx}
\usepackage{balance}
\usepackage{color}
\usepackage{comment}
\usepackage{csquotes}
\usepackage{amsmath,amssymb}
\usepackage{url}
\usepackage{hyperref}

\def\BibTeX{{\rm B\kern-.05em{\sc i\kern-.025em b}\kern-.08em
    T\kern-.1667em\lower.7ex\hbox{E}\kern-.125emX}}
    
\def\baselinestretch{0.955}

\newcommand{\EN}{{EN}\xspace}
\newcommand{\ENnospace}{{EN}}

\newcommand{\ENs}{{ENs}\xspace }

\newcommand{\eg}{{\em e.g.,}\ }
\newcommand{\ie}{{\em i.e.,}\ }
\newcommand{\sol}{{\em Reservoir}\xspace}

\begin{document}

\makeatletter
\def\ps@IEEEtitlepagestyle{%
  \def\@oddfoot{\mycopyrightnotice}%
  \def\@evenfoot{}%
}
\def\mycopyrightnotice{%
  {\footnotesize \textcolor{red}{\begin{tabular}[t]{@{}l@{}} This paper has been accepted for publication by the 20th International Conference on Pervasive Computing and Communications (PerCom 2022). © 2022 IEEE. \\ Personal use of this material is permitted. Permission from IEEE must be obtained for all other uses, in any current or future media, including reprinting/republishing \\ this material for advertising or promotional purposes, creating new collective works, for resale or redistribution to servers or lists, or reuse of any copyrighted \\ component of this work in other works.\end{tabular}}}
  \gdef\mycopyrightnotice{}
}


\title{\sol: Named Data for Pervasive Computation Reuse at the Network Edge}

\author{
\IEEEauthorblockN{Md Washik Al Azad}
\IEEEauthorblockA{University of Nebraska at Omaha\\
malazad@unomaha.edu}
\and
\IEEEauthorblockN{Spyridon Mastorakis}
\IEEEauthorblockA{University of Nebraska at Omaha \\
smastorakis@unomaha.edu}
}



\makeatletter
\patchcmd{\@maketitle}
  {\addvspace{0.5\baselineskip}\egroup}
  {\addvspace{-1.1\baselineskip}\egroup}
  {}
  {}
\makeatother

\maketitle

\begin{abstract}
In edge computing use cases (\eg smart cities), where several users and devices may be in close proximity to each other, computational tasks with similar input data for the same services (\eg image or video annotation) may be offloaded to the edge. The execution of such tasks often yields the same results (output) and thus duplicate (redundant) computation. Based on this observation, prior work has advocated for \enquote{computation reuse}, a paradigm where the results of previously executed tasks are stored at the edge and are reused to satisfy incoming tasks with similar input data, instead of executing these incoming tasks from scratch. 
However, realizing computation reuse in practical edge computing deployments, where services may be offered by multiple (distributed) edge nodes (servers) for scalability and fault tolerance, is still largely unexplored. To tackle this challenge, in this paper, we present \sol, a framework to enable pervasive computation reuse at the edge, while imposing marginal overheads on user devices and the operation of the edge network infrastructure. \sol takes advantage of Locality Sensitive Hashing (LSH) and runs on top of Named-Data Networking (NDN), extending the NDN architecture for the realization of the computation reuse semantics in the network. Our evaluation demonstrated that \sol can reuse computation with up to an almost perfect accuracy, achieving 4.25-21.34$\times$ lower task completion times compared to cases without computation reuse.
\end{abstract}

\vspace{-0.3cm}


\maketitle

\section{Introduction}

Edge computing has emerged as a paradigm that offers low-latency access to computing resources for computation-intensive, latency-sensitive applications (\eg Augmented Reality)~\cite{shi2016edge}. Edge computing use cases (\eg smart cities) may involve several users and devices in close proximity that offload tasks with similar input data for the same services (\eg image or video annotation) to computing resources at the network edge~\cite{guo2018foggycache}. The execution of tasks for the same service(s) and with similar input data often yields the same results (output), thus resulting in duplicate (redundant) computation~\cite{guo2018foggycache, lee2019case}. This observation enables the reuse of computation, so that the results of executed tasks are stored by edge nodes (servers) in order to be reused and satisfy \enquote{similar} incoming tasks (\ie tasks for the same service(s) with similar input data) in the future, instead of executing these incoming tasks from scratch. 

Prior work has explored computation reuse, showing that it can substantially reduce the execution of duplicate computation and speed up the execution of tasks at the edge~\cite{guo2018foggycache, lee2019case, guo2018potluck}. 
The vast majority of prior work has assumed that similar tasks will be offloaded to the same Edge Node (\EN), so that the results of previously executed similar tasks can be reused. However, the networking mechanisms to enable computation reuse in practical edge computing deployments, where each service may be offered by multiple \ENs in a distributed manner for fault tolerance, scalability, and load-balancing purposes, are still largely unexplored. To offer computation reuse in such deployments, the edge network infrastructure needs to be aware of the computation reuse semantics in order to identify similar tasks and forward them to the same \ENnospace(s). At the same time, given the potential resource-constrained nature of user devices as well as the strict low-latency requirements and potentially critical nature of applications, the realization of computation reuse must: (i) offer high confidence about the accuracy of the reused results (\ie the reused results must be the same as the results of the execution of an incoming task from scratch); and (ii) impose marginal overheads on user devices and the operation of the edge network infrastructure.

To tackle these challenges, in this paper, we present \sol\footnote{In the same manner as a natural or artificial lake that stores water for future supply, \ENs in \sol store the execution results of tasks in order to reuse them and satisfy incoming tasks with similar input data for the same service(s) in the future without executing these incoming tasks from scratch.}, a framework that aims to achieve pervasive computation reuse at the network edge. \sol runs on top of Named-Data Networking (NDN)~\cite{zhang2014named}, capitalizing on the data-centric nature and semantically meaningful naming of the NDN communication model. At the same time, \sol extends the NDN architecture and takes advantage of Locality Sensitive Hashing (LSH) mechanisms~\cite{lv2007multi, andoni2015practical} for the realization of the computation reuse semantics directly in the edge network infrastructure. The contributions of our work are the following:

\begin{itemize}[wide, labelwidth=!, labelindent=0pt]

\item We present the \sol design, which features unified mechanisms for task naming, the light-weight identification of similar tasks in the network, and the forwarding of such tasks towards the same \ENs, so that applications can capitalize on the benefits of computation reuse.

\item We implement a \sol prototype, which we evaluate through a study that involves the assessment of its components in isolation, real-world experiments of the \sol design as a whole, and a network simulation study of the \sol prototype. Our evaluation results demonstrate that \sol incurs marginal performance overheads per offloaded task on user devices (less than 1.7ms) and the network forwarding operation (less than 5$\mu$s). \sol also reuses computation with an almost perfect accuracy (up to 100\% in certain cases), achieving 4.25-21.34$\times$ lower task completion times compared to cases where computation reuse is not applied.

\end{itemize}

The rest of our paper is organized as follows: in Section~\ref{subsec:background}, we give a brief background on NDN and LSH, and present prior related work. In Section~\ref{sec:overview}, we highlight application use cases that can benefit from computation reuse and we motivate the \sol framework. In Section~\ref{sec:design}, we present the design of \sol, and, in Section~\ref{sec:eval}, we present the evaluation of \sol. 
Finally, in Section~\ref{sec:concl}, we conclude our paper.

\section{Background and Prior Related Work}
\label{subsec:background}

\noindent\textbf{Named-Data Networking:} Named-Data Networking (NDN)~\cite{zhang2014named} features a receiver-driven, name-based communication model. In NDN, data consumer applications send requests for data, called \emph{Interest packets} (or Interests for short). Interests based on their names are forwarded by the NDN network towards data producer applications, which respond with \emph{Data packets}. Each Data packet contains the requested data and is cryptograhically signed by its producer, while it also carries the producer's signature at rest and in transit across the network.

The NDN network consists of NDN forwarders~\cite{nfd-dev}, which forward Interest packets based on their names from consumers towards producers. To achieve that, each NDN forwarder is equipped with three main data structures: (i) a Forwarding Information Based (FIB), which contains entries of name prefixes along with one or more outgoing interfaces, and it is used for Interest forwarding purposes; (ii) a Pending Interest Table (PIT), which contains Interests that have been recently forwarded, but have yet to retrieve the corresponding Data packets; and (iii) a Content Store (CS), which caches retrieved Data packets to satisfy future Interests for data with the same name. Each device (\eg server, mobile phone) running a consumer or a producer application may also run an NDN forwarder instance locally, which enables communication over NDN. The NDN network is stateful: each Interest leaves state in the PIT of NDN forwarders, while a Data packet satisfies the state left by the corresponding Interest, following the reverse network path of the corresponding Interest back to the requesting consumer(s). If an Interest $I_1$ is received by an NDN forwarder while an Interest $I_2$ with the same name currently exists in the forwarder's PIT (pending Interest), $I_1$ will be aggregated with the existing PIT entry and will not be forwarded towards the producer. Once the requested Data packet is received by the forwarder, this Data packet will satisfy both $I_1$ and $I_2$. Finally, this Data packet will be forwarded back to all requesting consumers.

\noindent\textbf{Locality Sensitive Hashing:} Locality Sensitive Hashing (LSH) is an algorithmic technique, where similar data items are hashed with high probability in the same bucket(s) of a hash table~\cite{indyk1998approximate}. LSH is used as a mechanism to search for the k-nearest neighbors of an incoming item $i$ by applying a hash function $h$ to $i$ and using the resulting hash $h(i)$ as the index of the bucket(s) of a hash table to be searched for the nearest neighbors of $i$. To improve the accuracy of the search process, given the high-dimensionality of certain item types (\eg images, videos), a family of hash functions $h_1, h_2, ..., h_n$ may be applied to incoming items, where the resulting hash of each function is used as the index of a different hash table~\cite{indyk1998approximate, gionis1999similarity, datar2004locality}. However, such approaches may require the maintenance of a large number of hash tables (up to more than a hundred hash tables)~\cite{haghani2008lsh}. To avoid the need for maintaining large numbers of hash tables and make the use of LSH practical, multi-probe LSH approaches have been proposed~\cite{lv2007multi, andoni2015practical}. In multi-probe LSH, a small number of hash tables needs to be maintained, while multiple buckets of each table are \enquote{probed} (searched), which are likely to contain similar items.


\noindent\textbf{Computation reuse in edge computing research:} The concept of computation reuse has been recently explored to speed up the execution of tasks at the edge. Lee \textit{et al.} performed an empirical evaluation of applying computation reuse in the context of different edge computing applications, such as face recognition and matrix multiplication~\cite{lee2019case}. Guo \textit{et al.} proposed Potluck to reuse the results of tasks with similar inputs across different applications running on the same device~\cite{guo2018potluck}, while they also proposed FoggyCache to achieve the reuse of computational task results across different devices~\cite{guo2018foggycache}. In addition, Drolia \textit{et al.} proposed Cachier, a system that uses edge as a specialized cache for recognition applications and applies optimization techniques to minimize task execution times, such as leveraging the spatiotemporal locality of tasks, offline analysis of edge services/applications, and online estimates of network conditions~\cite{drolia2017cachier}. Recently, Meng \textit{et al.} designed Coterie to take advantage of the similarity among background environment frames in multi-player Virtual Reality and proposed a technique to increase the similarity among these frames~\cite{meng2020coterie}. In NDN, Mastorakis \textit{et al.} proposed ICedge, an edge computing framework for the adaptive offloading and forwarding of tasks towards edge computing resources~\cite{mastorakis2020icedge}. In addition, ICedge proposed a preliminary design to facilitate the reuse of computation through NDN naming and forwarding.

\begin{table}[t]
\caption{Design properties of \sol in comparison to prior related work (Y: Yes, N: No, L: Limited).}
\resizebox{\columnwidth}{!}{%
\label{Table:relwork}
\begin{tabular}{|c|c|c|c|c|c|c|}
\hline
\textbf{}                                                                                  & Cachier & Potluck & FoggyCache & \multicolumn{1}{l|}{Coterie} & \multicolumn{1}{l|}{ICedge} & \multicolumn{1}{l|}{\textit{\textbf{Reservoir}}} \\ \hline
\begin{tabular}[c]{@{}c@{}}Reuse at \\ user devices\end{tabular}                                                                      & N       & Y       & N          & Y                            & L                           & \textit{\textbf{Y}}                              \\ \hline
\begin{tabular}[c]{@{}c@{}}Reuse in \\ edge network\end{tabular}                                                                       & N       & N       & N          & N                            & L                           & \textit{\textbf{Y}}                              \\ \hline
Reuse at ENs                                                                               & L       & N       & Y          & N                            & Y                           & \textit{\textbf{Y}}                              \\ \hline
\begin{tabular}[c]{@{}c@{}}Supports practical \\ edge computing \\ deployments\end{tabular} & N       & N       & N          & N                            & L                           & \textit{\textbf{Y}}                              \\ \hline
\end{tabular}
}
\vspace{-0.5cm}
\end{table}

\noindent \textbf{How does \sol differ from prior work?} As we summarize in Table~\ref{Table:relwork}, the vast majority of prior work: (i) achieves reuse only at user devices or \ENs; and (ii) assumes that each service will be offered by a single \EN or that similar tasks will be offloaded to the same \ENnospace(s), so that the results of previously executed tasks can be reused. However, in practical edge computing deployments, a service may be offered by multiple \ENs in a distributed fashion for scalability, fault tolerance, and load balancing purposes. Systems, such as Potluck and Cachier, have primarily focused on cache optimization mechanisms, while systems such as Coterie are application-specific. ICedge offers the means for the adaptive forwarding of tasks towards multiple \ENs, so that computation reuse is facilitated. However, in ICedge, NDN forwarders need to process tasks offloaded by each application in a different way, which needs to be determined by application developers. This makes task forwarding complicated, incurring task processing/forwarding delays. At the same time, ICedge does not utilize inherent features of the NDN architecture, such as in-network caching, so that similar tasks can reuse results that may be cached in the edge network, without the need to be forwarded at all to \ENs. To this end, \sol proposes a universal design that can be used by all applications at the network edge, exploiting inherent features of the NDN architecture and extending the NDN architectural design. \sol includes unified mechanisms for task naming, the light-weight identification of similar tasks in the network, and the forwarding of such tasks towards the same \ENs, so that applications can capitalize on the benefits of computation reuse, while incurring marginal overheads on user devices and the network performance.

\section{Use Cases and Motivation}
\label{sec:overview}


\subsection{Computation Reuse Use Cases}

To highlight the need for computation reuse, we present three use cases below. 

\begin{figure}[!t]
 \centering
 \includegraphics[width=0.85\columnwidth]{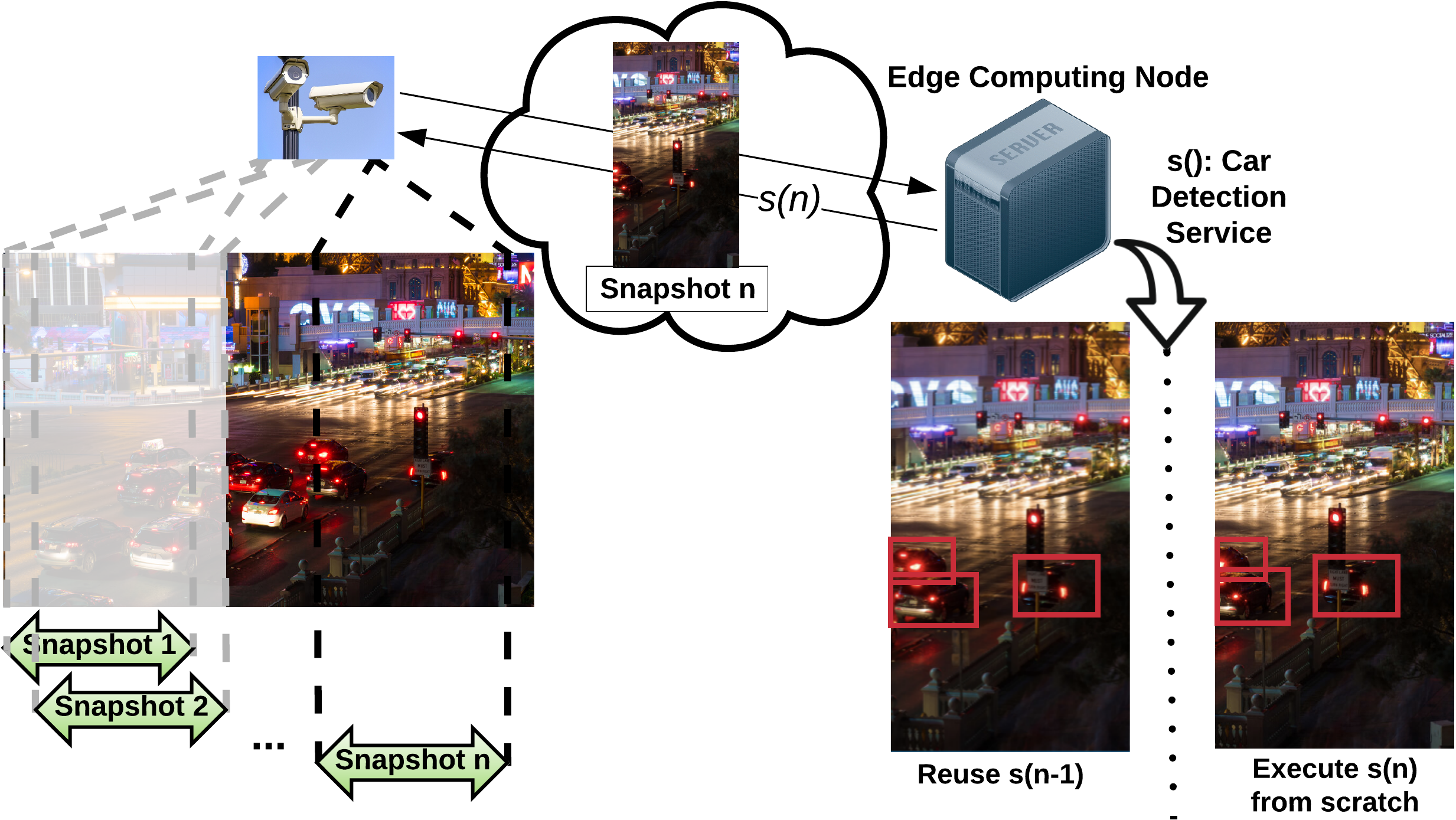}
 \vspace{-0.2cm}
 \caption{A video surveillance use case, where traffic snapshots are captured in a smart city for traffic management purposes.} 
 \label{Figure:re-use}
 \vspace{-0.5cm}
\end{figure}

\noindent \textbf{Traffic monitoring in a smart city:} Let us consider a video surveillance use case (Figure~\ref{Figure:re-use}), where cameras capture traffic snapshots in a smart city for traffic management purposes. Such cameras may have snapshot (frame) capture rates ranging from a snapshot every couple of seconds up to several snapshots per second, while they may also rotate to capture snapshots across their full capture width. For example, a camera in Figure~\ref{Figure:re-use} may rotate to estimate the traffic volume for all the lanes travelling in the same direction of an intersection, Consecutive camera snapshots, which may be considerably similar, are offloaded to an \EN, so that the number of cars is detected and the traffic volume is estimated. As a result, executing a car detection service on consecutive snapshots (\eg snapshots $n-1$ and $n$) may yield the same execution results. To this end, there is no need to execute the service from scratch for snapshot $n$, but we can rather reuse the results of service execution for snapshot $n-1$ (\ie the previous snapshot).  Even if the camera is stable (does not rotate), with capture rates 
up to several snapshots per second, it is likely that consecutive snapshots will still be considerably similar. In addition to having an \EN reusing the execution results of previously offloaded snapshots, the camera itself can cache the results of previous snapshots (if its resources allow), so that they are reused locally for consecutive (similar) snapshots, without the need to offload these snapshots at the edge.

\noindent \textbf{Cognitive assistance application:} 
Let us consider a cognitive assistance application, which enables visual search by identifying objects in scenes captured by Augmented Reality (AR) headsets or pictures taken by mobile phones. For example, users exploring a city may capture scenes of famous sights with their AR headsets or take pictures of sights with their mobile phones to acquire more information/content about these sights, such as podcasts, videos, and the sights' history. In this context, 
users exploring New York may capture scenes or take pictures of the Statue of Liberty (\eg from different angles or distances) and offload these scenes/pictures as the inputs of tasks for a service at the edge, which identifies the sight and returns information/content about it. These scenes/pictures are similar, thus yielding the same execution results.

\noindent \textbf{Smart homes equipped with Internet of Things (IoT):} Let us consider a virtual assistant application in smart homes equipped with IoT devices. Such an application may require speech identification, so that residents can control the IoT devices through voice commands. Residents of the same or nearby smart homes may invoke semantically similar commands that result in the same action (\eg turning on the lights in a room). The processing output among semantically similar commands can be reused/shared without the need to repeatedly run a speech identification service at the edge for all the semantically similar commands that residents invoke.

In all previous use cases, a massive number of devices may offload large numbers of similar tasks at the edge. For example, certain cities in China are estimated to have up to 2.5M CCTV cameras each~\cite{cctv-countries}, while major cities in Europe and the US are also estimated to have several tens or even hundreds of thousands of CCTV cameras~\cite{cctv-countries, cctv-london}. Popular sights around the world attract several millions of visitors annually and up to tens of thousands of visitors daily. For example, the Statue of Liberty attracts about 4.5M visitors annually and about 10K visitors daily~\cite{liberty}, while the Louvre museum attracts 10M visitors annually and about 15K visitors per day, containing more than 380K objects and displaying 35K works of art~\cite{louvre}. At the same time, the number of IoT devices is expected to reach 75 billions by 2025~\cite{cisco}. Such use cases may result in the execution of massive amounts of duplicate (redundant) computation, since offloaded tasks may include similar inputs. To this end, computation reuse can result in: (i) eliminating the execution of duplicate computation; (ii) reducing the execution times of tasks; and (iii) reducing the usage of computing resources at the edge and ensuring their effective utilization. 

We acknowledge that computation reuse may have security and privacy implications, which deserve further investigation. However, in this paper, we do not focus on such implications, but we rather take a first step to investigate whether computation reuse as a concept can be achieved in practical edge computing deployments. We further acknowledge that computation reuse may be effective in use cases, where temporal, spatial, or semantic correlation exists between the input data of offloaded tasks. However, not all use cases/applications exhibit the same degrees of correlation between the input data of their offloaded tasks (or they might exhibit minor correlations overall). There can also be use cases/applications, where minor differences in the input data may yield different execution results (\eg cryptographic operations). \sol can accommodate task input data with different degrees of correlation (as we present in Section~\ref{sec:eval}), while it can also accommodate (as we discuss in Section~\ref{sec:design}): (i) use cases with negligible correlation between task input data; and (ii) use cases, where minor differences in the input data may yield different execution results. 


\subsection {\sol Motivation}

Let us consider the use cases we described above in the context of a practical edge computing deployment, where services may be offered by multiple available \ENs. In the traffic monitoring use case, the execution results for similar snapshots may be cached and reused directly by a CCTV camera. 
However, this may not be possible, since CCTV cameras have limited cache/storage resources, thus being able to store up to a certain number of tasks. In this case, unless similar snapshots are forwarded by the edge network towards the same \EN, the reuse of computation may not be possible. For example, in the scenario of Figure~\ref{Figure:re-use}, if snapshot $n-1$ is forwarded to \ENnospace$_1$, while snapshot $n$ is forwarded to \ENnospace$_2$ (\ENnospace$_1$ and \ENnospace$_2$ both offer the car detection service), reuse will not be possible. The same applies to the example of the cognitive assistance application, where pictures depicting the same sight taken by different visitors need to be forwarded towards the same \EN for computation reuse to be possible, and the virtual assistant application, where contextually similar voice commands invoked by residents of neighboring smart homes need to be forwarded towards the same \EN. To this end, a framework like \sol is needed to achieve pervasive computation reuse on behalf of applications in such practical edge computing deployments.

\noindent \textbf{Why do we realize \sol over NDN:} NDN offers a communication model that makes the network directly aware of the communication context, since packets carry semantically meaningful names. \sol builds on top of and extends NDN naming and forwarding combined with LSH to: (i) make the network aware of the computation reuse semantics; (ii) enable the identification of similar tasks in the network in a light-weight manner; and (iii) enable the forwarding of tasks in a \enquote{computation reuse aware} fashion with nominal performance overhead, so that similar tasks are forwarded towards the same \ENs, facilitating the reuse of computation.

\section{\sol Design}
\label{sec:design}


\subsection{Design Assumptions, Overview, and Goals}
\label{subsec:overview}

We assume the existence of edge networks (\enquote{cloudlets})~\cite{satyanarayanan2009case}, which consist of \ENs and NDN forwarders. The \ENs are server-class nodes with computing and storage resources, and offer a set of services (\eg object recognition, face detection) to users. \ENs may either be a single hop (\eg directly attached to LTE/5G base stations) or a few hops away from users (\eg at the edge of the core network). Users offload tasks, which are forwarded by NDN forwarders towards the \ENs of an edge network. Each task invokes one of the offered services, while it may also carry input data to be passed to the invoked service for its execution (\eg an object detection service may expect an image as input data). Each \EN and user device have a name prefix, which is used for direct communication with this specific \EN or device.

\begin{figure}[t]
 \centering
 \includegraphics[width=1\columnwidth]{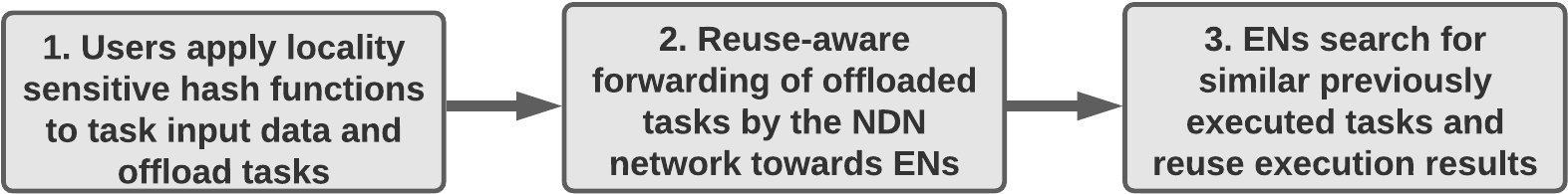}
 \vspace{-0.5cm}
 \caption{Workflow of the \sol operation.}
 \label{Figure:workflow}
 \vspace{-0.5cm}
\end{figure}

The workflow of the \sol operation is illustrated in Figure~\ref{Figure:workflow}. In \sol, a task is represented as a request (Interest packet) that identifies the service to be invoked in its name. First, users apply locality sensitive hash functions to the input data of a task and attach the resulting hash to the name of the Interest. 
As a result, tasks that invoke the same service with similar inputs are likely to have identical names, thus being able to: (i) reuse results of similar previously executed tasks that might be cached in the CS of NDN forwarders;
and (ii) get aggregated in the network 
if a similar task is currently pending in the PIT of NDN forwarders. 
Each user device also runs a local NDN forwarder instance, thus it can cache the results of previously executed tasks for reuse in its local CS and aggregate subsequent similar tasks in its local PIT depending on the availability of its resources.
If (i) and (ii) are not possible, offloaded tasks will be forwarded towards \ENs by NDN forwarders in a \enquote{computation-reuse aware} manner, so that similar tasks are forwarded towards the same \EN to facilitate the reuse of the execution results among similar tasks. To make NDN forwarders aware of the semantics of computation reuse, we extend their Interest forwarding pipeline by introducing a reuse FIB (rFIB) data structure (Figure~\ref{Figure:forwarding}), which stores reuse-related information in the network. Finally, once a task $t_{new}$ is received by an \EN, LSH will enable the \EN to effectively search for a previously executed task $t_{previous}$ of the same service, so that the similarity between the input data of $t_{new}$ and $t_{previous}$ exceeds a certain similarity threshold. If such a task $t_{previous}$ is found, the \EN will reuse the results of $t_{previous}$, otherwise, the \EN will execute $t_{new}$ from scratch.

Overall, the \textbf{goal} of \sol is not only to facilitate the reuse of computation at the network edge, but also to: 

\begin{itemize}[wide, labelwidth=!, labelindent=0pt]

\item \textbf{Impose nominal performance overhead on user devices} for receiving the benefits of computation reuse.

\item \textbf{Impose nominal performance overhead on the NDN network} for the identification and forwarding of similar tasks in a reuse aware manner.

\item \textbf{Enable \ENs to find} the execution results of previously executed similar tasks \textbf{as quickly as possible.}

\end{itemize}

\subsection{Task Namespace Design}
\label{subsec:namespace}

As we mentioned in Section~\ref{subsec:overview}, a task is represented as an Interest packet. Each task has a name \enquote{/<service-name>/task/<hash-of-input>}. The first name component identifies the service to be invoked by a task, the second component is the keyword \enquote{task}, while the third component is the resulting hash after applying a locality sensitive hash function to the task input data. If a family of such hash functions needs to be applied to the input data, the resulting hashes will be concatenated and their concatenation will be attached as the task's third name component. For example, a task with a name \enquote{/OpenPose/task/6E810F} invokes the OpenPose service~\cite{cao2018openpose} with 6E810F being the hash of the task's input data. This Interest may additionally carry in its \enquote{application parameters} field~\cite{ndn-parameters}: (i) a deadline $\delta$ indicating the maximum latency that a task can tolerate until its execution results are returned to users; (ii) a similarity threshold, which the similarity between the input data of the incoming task $t_{new}$ and a previously executed task $t_{previous}$ for the same service needs to exceed, so that the results of $t_{previous}$ are reused for the execution of $t_{new}$. This threshold can be determined by applications depending on their requirements (we further discuss that in Section~\ref{subsec:reuse}); and (iii) the actual task input data to be passed to the invoked service. For input data of large sizes that cannot fit into a single Interest packet, we present a more sophisticated input passing mechanism in Section~\ref{subsec:offloading}. 

This namespace enables \sol to capitalize on the NDN architectural features of in-network caching and Interest aggregation, thus limiting the number of tasks that are forwarded to \ENs and the load of \ENs. \sol is able to: (i) retrieve execution results for similar tasks that may be cached in the NDN network or user devices, 
so that such tasks may not need to be forwarded to \ENs for reuse or execution from scratch; and (ii) aggregate similar tasks that are simultaneously offloaded in the NDN network or on user devices, so that a single task, among all the similar tasks, is forwarded to an \EN, while the execution results are returned to all requesting users/devices. These features are a consequence of the fact that applying locality sensitive hash functions to similar data is likely to result in the same hashes, so that tasks for the same service and similar input data may have the same names. 

\subsection{Task Offloading Mechanisms}
\label{subsec:offloading}

\begin{figure}[t]
	\centering
	\begin{subfigure}{.295\columnwidth}
		\centering
		\includegraphics[scale=0.26]{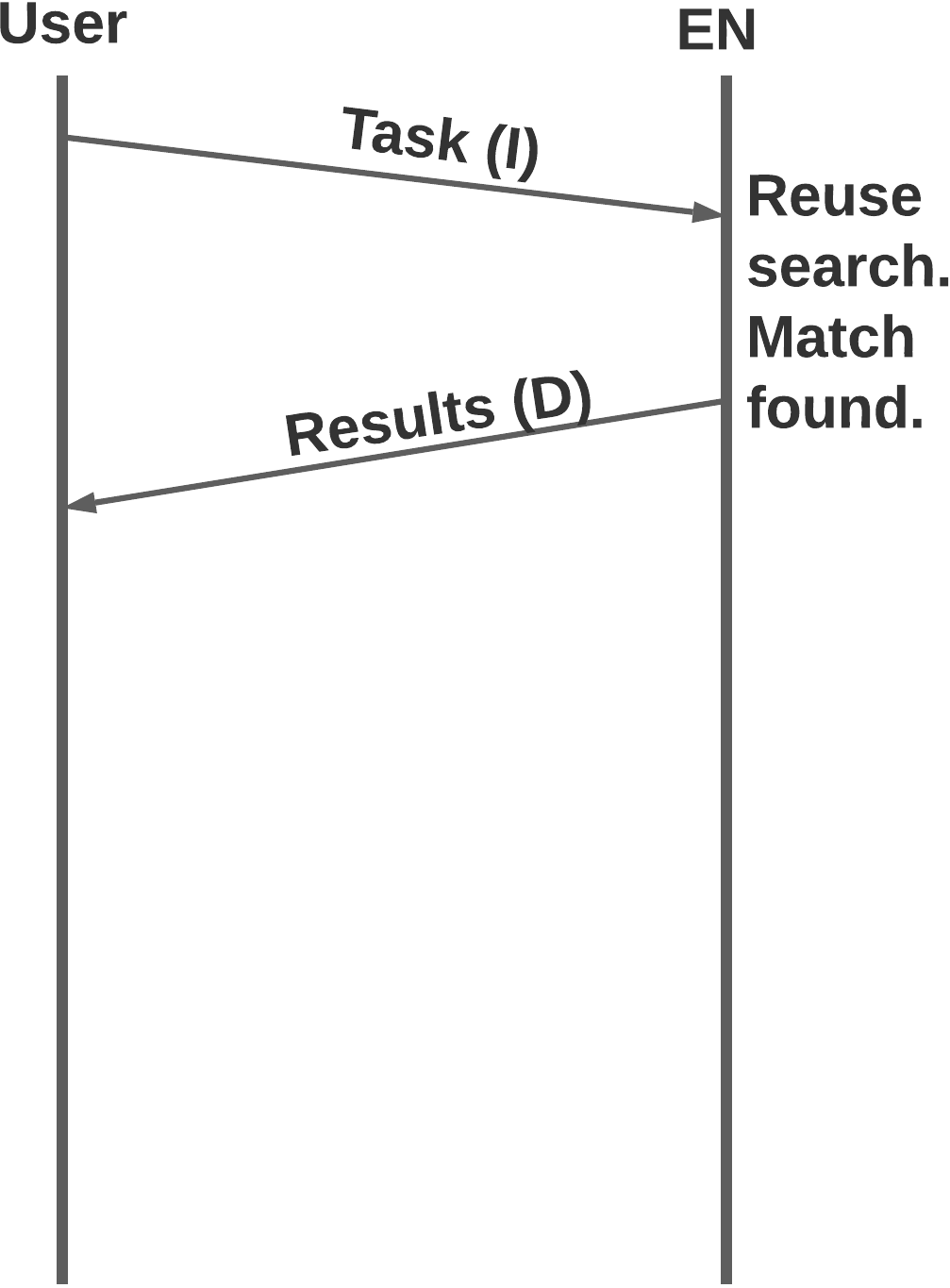}
		\caption{Computation reuse at the \EN. \\}
		\label{Figure:scenario1}
	\end{subfigure} \hfill
	\begin{subfigure}{.335\columnwidth}
		\centering
		\includegraphics[scale=0.26]{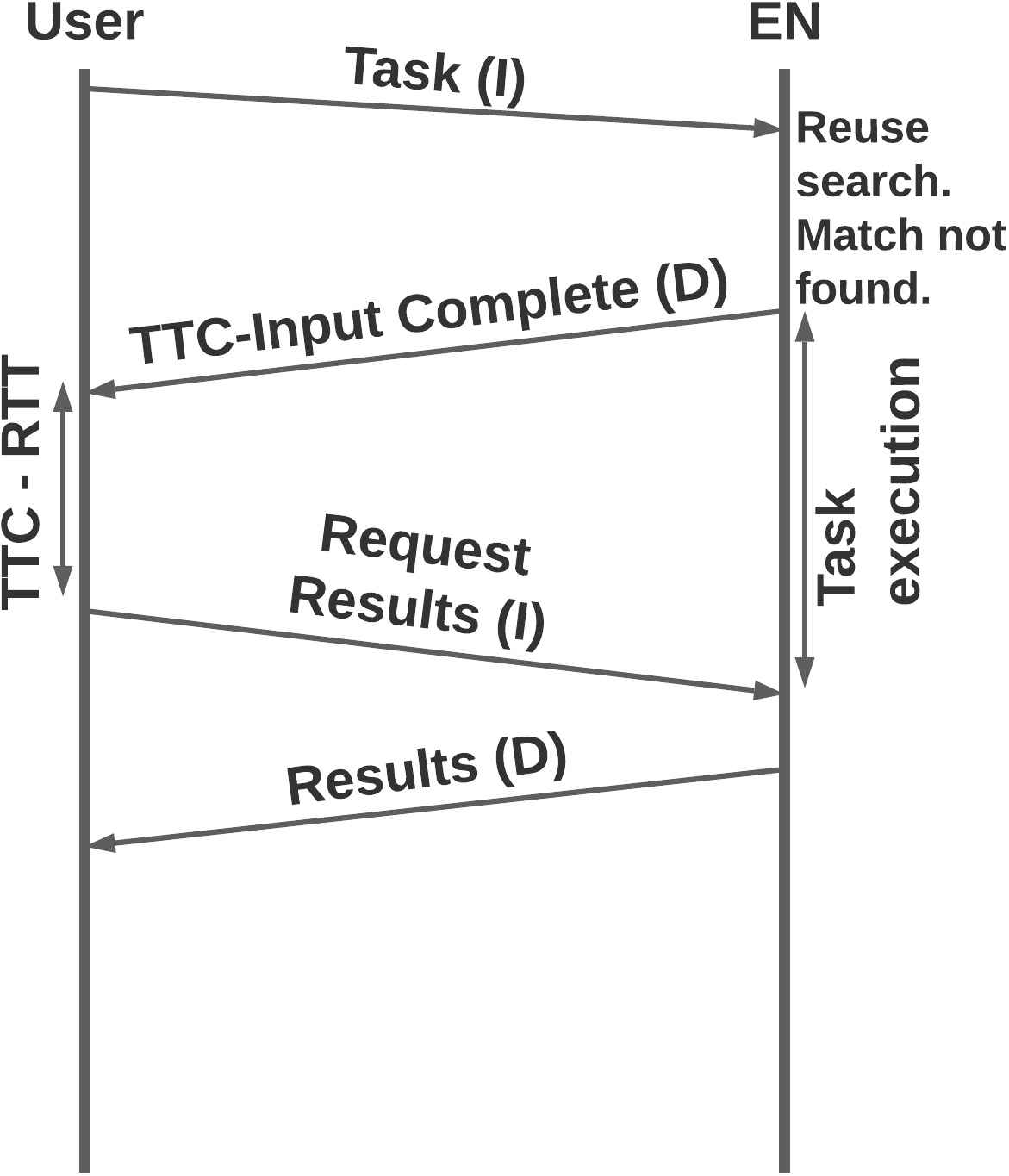}
		\caption{Execution from scratch at the \EN. \\}
		\label{Figure:scenario2}
	\end{subfigure} \hfill
	\begin{subfigure}{.34\columnwidth}
		\centering
		\includegraphics[scale=0.26]{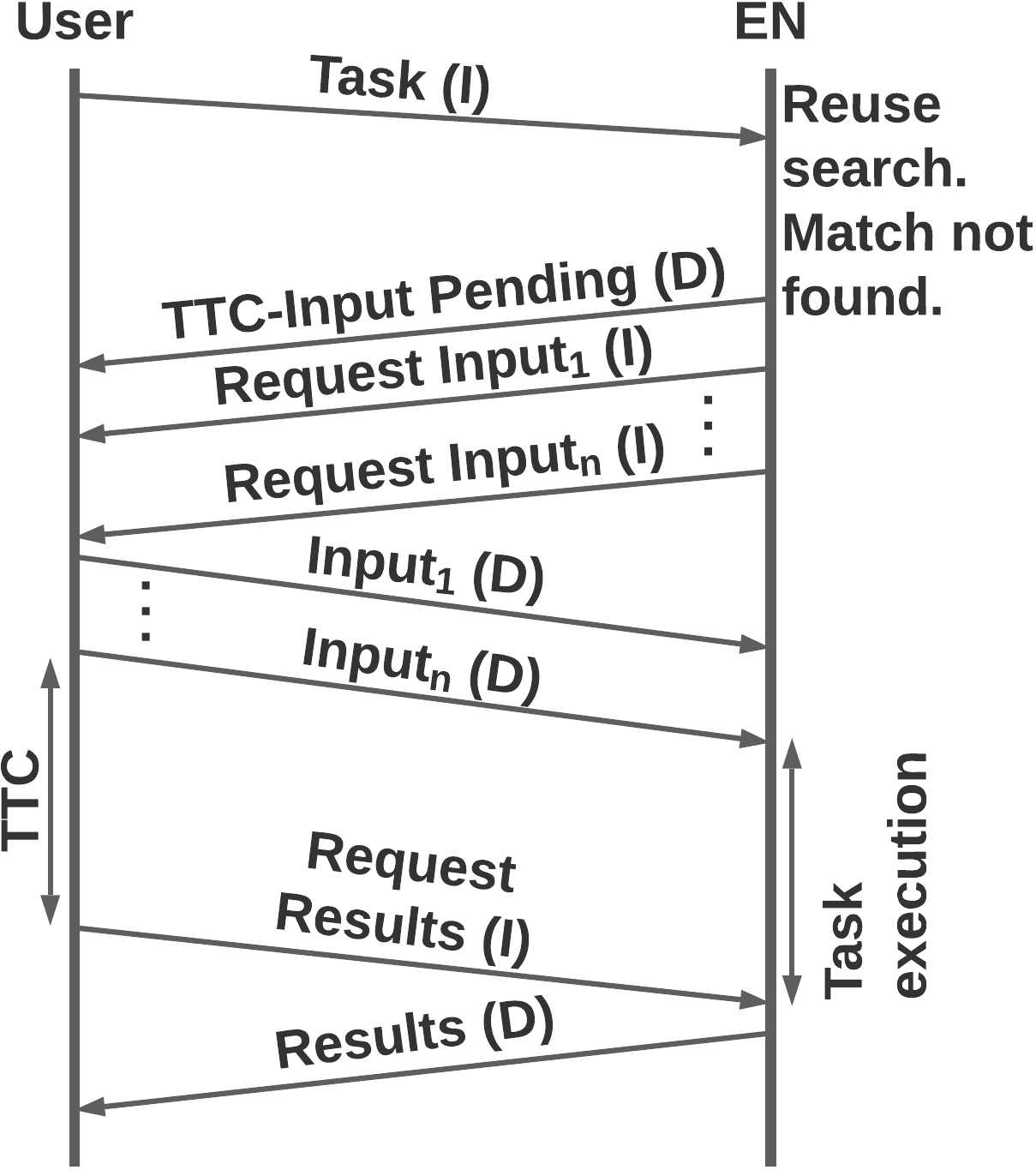}
		\caption{Execution from scratch at the \EN (large size of task input data).}
		\label{Figure:scenario3}
	\end{subfigure}
	\vspace{-0.2cm}
	\caption{\sol task offloading mechanism design (I: Interest packet, D: Data packet).}
	\label{Figure:offloading}
	\vspace{-0.5cm}
\end{figure}

We present the design of our task offloading mechanism in Figure~\ref{Figure:offloading} under different scenarios, where a task is forwarded to an \EN (no cached results are found in the network or user devices). In Figure~\ref{Figure:scenario1}, a task is offloaded by a user and is forwarded to an \EN by the NDN network. The task is represented as an Interest following the namespace of Section~\ref{subsec:namespace}. The \EN receives the task and determines whether it has stored results of previously executed similar tasks that can be reused (whether the similarity between the input of the incoming task and the input of previous tasks for the same service exceeds a certain threshold). The \EN is able to find a matching previous task and returns its results in a Data packet to the user.

In Figure~\ref{Figure:scenario2}, the \EN receives a task from a user and searches for a previous task to reuse, however, a task to be reused is not found. In this case, the \EN will respond with a Data packet containing: (i) an estimate of the amount of time needed for the execution of the task, called Time To Completion (TTC); and (ii) the name prefix of the \EN. To be able to estimate TTC, \ENs maintain statistics about the execution of the services over time. Based on this Interest-Data exchange, the user estimates the Round Trip Time (RTT) to the \EN and sends an Interest with a name \enquote{/<\EN-prefix/<service-name>/task/<hash-of-input>} to request the task execution results from the \EN after a time interval equal to TCC-RTT. The first component of this Interest's name refers to the name prefix of the \EN executing the task (ensuring that the Interest is forwarded to the right \EN), the second one to the invoked service, the third one to the keyword \enquote{task}, and the last one to the locality sensitive hash of the task input data. Once the \EN receives this Interest and the task execution results become available, the \EN sends the results back to the user.

In Figure~\ref{Figure:scenario3}, we present the offloading mechanism in cases of large task input data sizes. In such cases, the \EN requests the input data from the user by sending one or more Interests (depending on the input data size) to the user, so that the input data is passed to the \EN. To achieve that, the user attaches to the initial task sent to the \EN (\eg in the \enquote{application parameters} field of this Interest): (i) the estimated size of the input; and (ii) the name prefix of the user device. Subsequently, the \EN will use (i) to determine the number of Interests to be sent for the retrieval of the input data and (ii) to communicate with the user directly. Once the input is passed to the \EN, the task execution will begin and the user will retrieve the results once they are available in the same manner as in Figure~\ref{Figure:scenario2}. 



\noindent \textbf{Design tradeoffs:} The design of the offloading mechanism involves several tradeoffs. First, this mechanism adopts a pull-based model for passing the task input data from users to \ENs and for retrieving the task execution results from \ENs in accordance with the underlying pull-based NDN communication model. This may result in additional delay during the offloading process. In cases where this delay cannot be tolerated, the input data can be attached to the \enquote{application parameters} field of Interests and be sent (pushed) from the user to the \EN, while the execution results can be also attached to the \enquote{application parameters} field of an Interest and be sent from the \EN to the user as soon as they become available. This may also address issues related to the inaccurate estimation of TTC by \ENs, which we further discuss below, relaxing the need for accurate TTC estimates. Second, \ENs estimate TTC and users send Interests to retrieve the execution results right when the results become available in order to minimize the time that pending Interests (state) need to be maintained in the PIT of NDN forwarders. Alternatively, users could send an Interest for the execution results once the task execution begins, which will stay pending in the network until the results become available. This increases the amount of state that needs to be maintained by the network, but relaxes the need for accurate TTC estimates by \ENs. Third, in cases that the TTC estimation is not accurate, users may request the results before or considerably after they become available. If users request the results before they are available, \ENs can respond with an updated TTC estimation. If \ENs have the results earlier than the estimated TTC, they can proactively notify users about the earlier availability of the results. Finally, for tasks with large data inputs, the data inputs are passed to \ENs only if reuse is not possible. This reduces the burden on the network, since input data is only transferred to \ENs if necessary, however, it requires an extra RTT for passing the task input data to \ENs.


\subsection{Realizing Reuse Awareness in the Network}
\label{subsec:reuse_net}

\begin{figure}[!t]
 \centering
 \includegraphics[width=0.83\columnwidth]{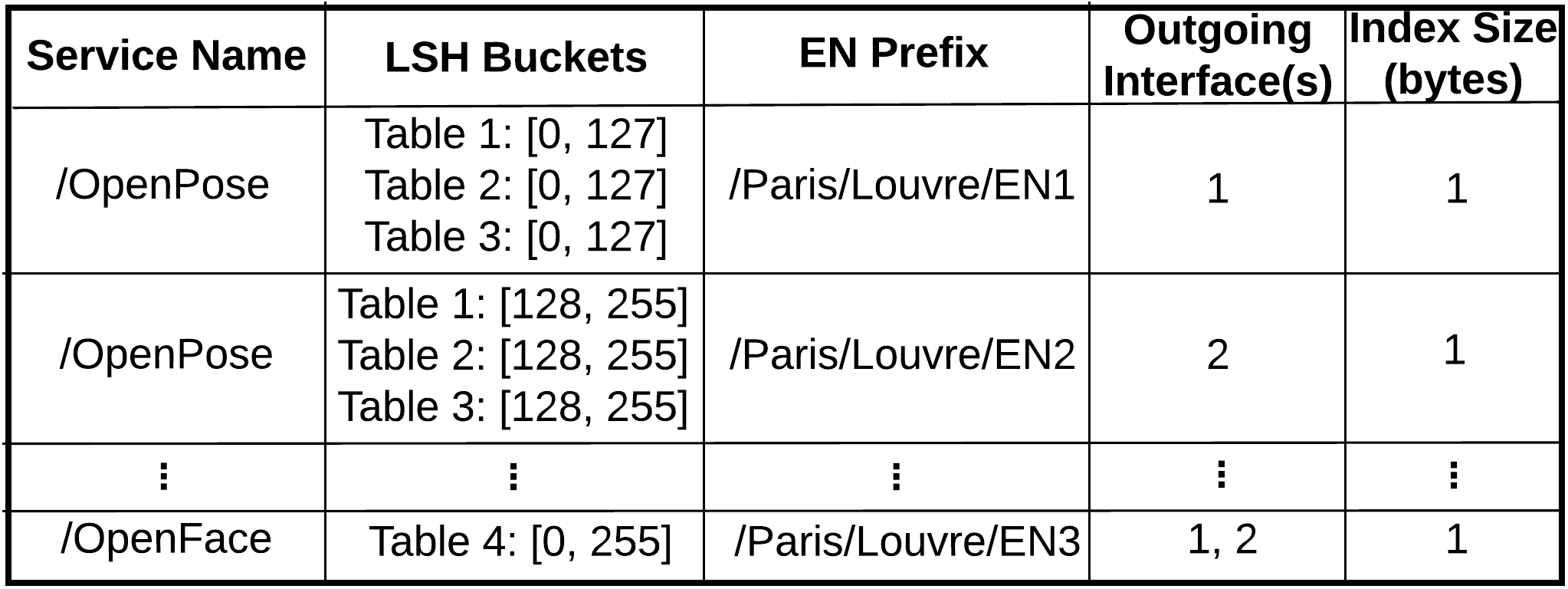}
 \vspace{-0.15cm}
 \caption{An example of the rFIB data structure.}
 \label{Figure:rFIB}
 \vspace{-0.4cm}
\end{figure}

\begin{figure}[!t]
 \centering
 \includegraphics[width=0.89\columnwidth]{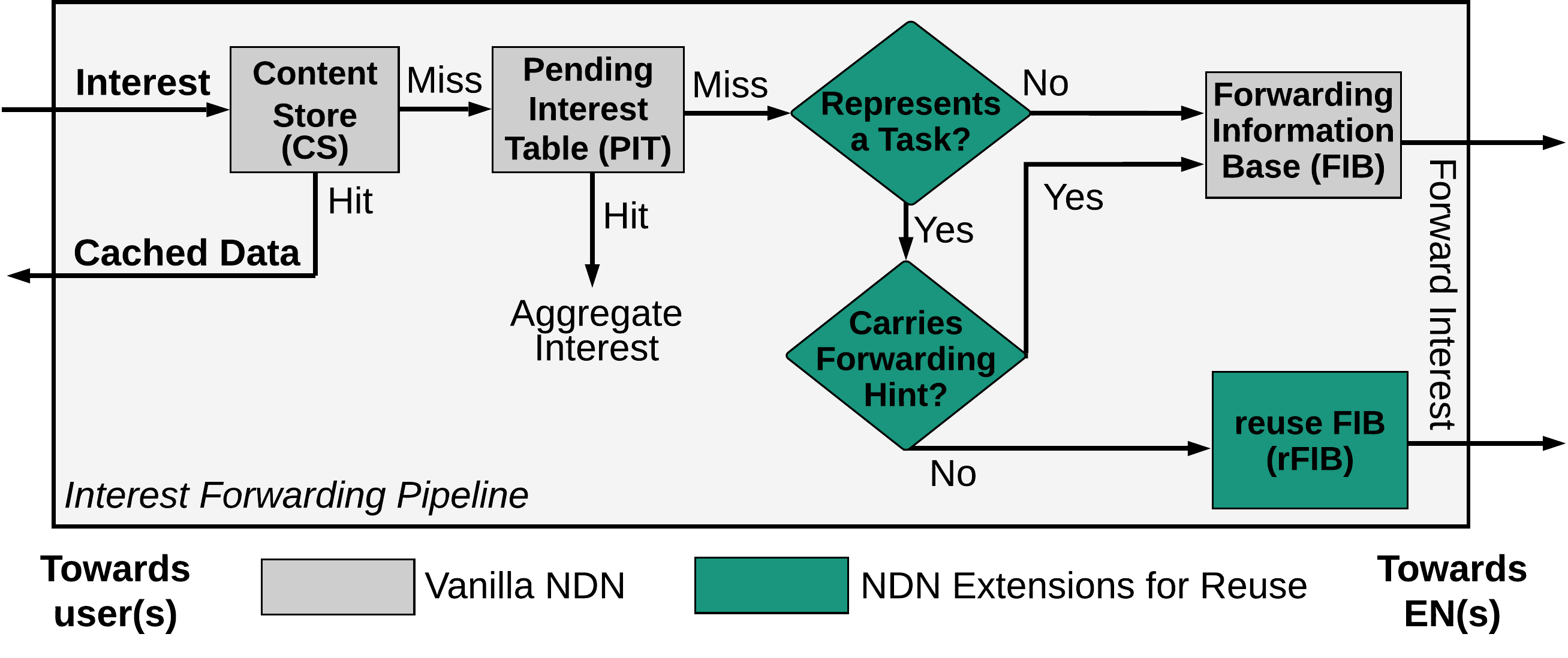}
 \vspace{-0.3cm}
 \caption{\sol Interest forwarding pipeline.}
 \label{Figure:forwarding}
 \vspace{-0.7cm}
\end{figure}

The edge network needs to be able to identify and forward similar tasks towards the same \EN, so that computation reuse is facilitated. To this end, NDN forwarders need to be aware of the semantics of computation reuse. This is achieved by introducing a reuse FIB (rFIB) data structure (Figure~\ref{Figure:rFIB}) and by extending the legacy NDN forwarding pipeline of Interests (Figure~\ref{Figure:forwarding}). rFIB contains entries that consist of: (i) the name of a service $s$; (ii) a range of LSH bucket indices that are handled by each \EN that offers $s$; (iii) the name prefix of the \ENnospace(s) offering $s$; (iv) the outgoing interface(s) towards each \EN offering $s$; and (v) the size (in bytes) of the index used by LSH (hash) tables, so that forwarders can extract the index of each table from the hash in the third component of the task name when multiple LSH tables are used for $s$. Given that similar data is highly likely to be hashed in adjacent, if not the same, buckets in multi-probe LSH~\cite{lv2007multi}, we aim to have ranges of consecutive bucket indices assigned to \ENs. This results in rFIB entries that also contain consecutive blocks of bucket indices, simplifying the overall rFIB lookup process. 

Once an Interest is received by an NDN forwarder, after performing a CS lookup for cached data and inserting the Interest into PIT, the forwarder checks if this Interest is a task (whether its second name component is the keyword \enquote{task}). If the Interest is not a task, the forwarder follows the legacy NDN forwarding pipeline by finding a match in FIB based on the Interest name and forwarding the Interest. If the Interest is a task, the forwarder will perform a lookup on rFIB. The forwarder will find the rFIB entries for the service specified in the first name component of the task name and select one of these entries based on the buckets handled by each \EN and the hash in the third component of the task name.

For example, based on the rFIB of Figure~\ref{Figure:rFIB}, a task with a name \enquote{/OpenPose/task/6E810F} indexes the 110th (6E in hex) bucket of table 1, the 129th (81 in hex) bucket of table 2, and the 15th (0F in hex) bucket of table 3. The indexed buckets of tables 1 and 3 are handled by \ENnospace1 of the Louvre museum in Paris, while the indexed bucket of table 2 is handled by \ENnospace2 of the Louvre. In such cases, the task will be forwarded to the \EN that handles the majority of the indexed buckets (\EN with name prefix \enquote{/Paris/Louvre/EN1} in our example) in order to maximize the chances of reusing a previously executed task. Subsequently, the forwarder will attach the name prefix of the matched \EN to the task as the task's forwarding hint~\cite{ndn-hint}. The forwarding hint is an additional name identifier that Interests can carry to indicate \enquote{where}, or at which \EN in our case (forwarding hint), to execute \enquote{which} task (Interest name). As a result, \textbf{the rFIB lookup will happen only once per task} by the first NDN forwarder in the edge network that receives the task to minimize the overhead on the task forwarding performance. If a task carries a forwarding hint (\ie an rFIB lookup has been previously performed for this task), the task will be forwarded based on its forwarding hint through a regular FIB lookup as all other Interests (Figure~\ref{Figure:forwarding}).

\subsection{Computation Reuse in the Edge Network or at the \ENs}
\label{subsec:reuse}

\vspace{-0.05cm}

In \sol, reuse of computation can happen either in the edge network or at the \ENs. In the edge network, NDN forwarders may cache the results of previously executed tasks, while LSH combined with our task namespace design 
enables the reuse (retrieval) of the results of similar tasks cached in the edge network. If a task is forwarded to an \EN, the LSH algorithm running on the \EN will use the hash included in the task's name (third name component) to find the bucket(s) of each hash table that are likely to have stored similar previously executed tasks. 
Subsequently, the \EN extracts the similarity threshold attached to the incoming task $t_{new}$ 
and returns a previously executed task $t_{previous}$ for the same service invoked by $t_{new}$, so that: (i) the input data of $t_{previous}$ has the highest similarity among stored previously executed tasks to the input data of $t_{new}$ (nearest neighbor of $t_{new})$; and (ii) the similarity between the input data of $t_{new}$ and $t_{previous}$ exceeds the similarity threshold. If a task $t_{previous}$ that satisfies these conditions cannot be found, the \EN executes $t_{new}$ from scratch following the mechanism of Section~\ref{subsec:offloading}. Note that \sol can support the use of various similarity forms and algorithms (\eg structural similarity, cosine similarity)~\cite{wang2004image, singhal2001modern}.


\noindent \textbf{Deciding on the proper similarity threshold value:} The similarity threshold value for each task impacts the extent that the results of previous tasks can be reused. However, its selection by applications provides flexibility. The value of this threshold depends on the nature of each application and the type/granularity of input data processing that each application needs. For example, an application that requires the detection of whether traffic exists in captured snapshots may find a lower similarity threshold adequate. On the other hand, an application that requires fine-grained processing of traffic snapshots (\eg detection of the exact number of vehicles) may need higher similarity thresholds to operate properly.


\noindent \textbf{Tasks with minor similarities between input data:} Not all use cases/applications involve tasks with input data, which exhibits temporal, spatial, or semantic similarity. In \sol, application developers can indicate whether they expect their applications to involve tasks with similar input data. \ENs can also identify applications that do not offload tasks with similar data over time. 
Applications with minor input data similarities can indicate to the edge network infrastructure that their tasks should not be treated in the context of computation reuse (by creating and offloading tasks with a special component in their names or a dedicated flag in their packet format). This enables: (i) applications to eliminate the overhead of generating a locality sensitive hash per task, thus generating a less expensive hash instead (\eg CRC32, SHA1), which will be used in the namespace of Section~\ref{subsec:namespace}; (ii) the edge network infrastructure to eliminate the overhead of performing \enquote{computation-reuse aware} task forwarding based on the rFIB, but rather forward tasks based on the FIB directly; and (iii) \ENs to eliminate the overhead of nearest neighbor searches. This mechanism also applies to applications where tasks with similar input data may yield substantially different results. 

\section{Evaluation}
\label{sec:eval}

In this section, we present the evaluation of \sol in four phases: (i) we evaluate components of the \sol design in isolation to study their performance and trade-offs; (ii) we conduct a real-world feasibility study of a \sol prototype; (iii) to scale up our experiments, we perform a network simulation study of the \sol prototype; and (iv) we compare \sol to ICedge~\cite{mastorakis2020icedge}, which offers a preliminary design to facilitate the reuse of computation.


\noindent \textbf{Task input datasets and edge services:} For our evaluation, we utilize real-world image datasets, which we describe in Table~\ref{tab:datasets}, as task input data. We use the FALCONN library~\cite{andoni2015practical} to realize LSH and perform hashing and similarity searches over these image datasets. In the rest of this section, we run each experiment ten times and we report on the average results.

\begin{table*}[!t]
\centering
\begin{minipage}{.7\linewidth}
\caption{Edge services and real-world datasets used for the evaluation of \sol.}
\vspace{-0.2cm}
\label{tab:datasets}
\resizebox{0.98\linewidth}{!}{%
\begin{tabular}{|c|c|c|c|c|c|}
\hline
\textbf{Dataset}                                              & \textbf{Type of data}                                                                                                                             & \textbf{\begin{tabular}[c]{@{}c@{}}Dataset \\ size\end{tabular}} & \textbf{\begin{tabular}[c]{@{}c@{}}Correlation \\ between data\end{tabular}} & \textbf{\begin{tabular}[c]{@{}c@{}}Performed data \\ processing (edge services)\end{tabular}}                                                                        & \textbf{\begin{tabular}[c]{@{}c@{}}Processing\\  granularity\end{tabular}} \\ \hline
MNIST~\cite{lecun1998mnist}                                                         & Images of handwritten digits                                                                                                                      & 70K                                                              & Low                                                                          & Digit identification                                                                            & Medium                                                             \\ \hline
Pandaset~\cite{pandaset}                                                      & \begin{tabular}[c]{@{}c@{}}Images captured by autonomous \\ vehicles in Silicon Valley, California\end{tabular}                                   & 49K                                                              & Low                                                                          & \begin{tabular}[c]{@{}c@{}}Detection of obstacles \\ and positions around \\ vehicles\end{tabular} & Fine                                                               \\ \hline
\begin{tabular}[c]{@{}c@{}}Stanford \\ Mobile AR~\cite{makar2013interframe}\end{tabular} & Images of different objects                                                                                                                       & 1K                                                               & Moderate                                                                     & Object identification                                                                           & Medium                                                             \\ \hline
CCTV1                                                         & \begin{tabular}[c]{@{}c@{}}Snapshots that we extracted from video \\ streams of CCTV cameras monitoring \\ traffic at intersections in the US\end{tabular} & 5K                                                               & High                                                                         & \begin{tabular}[c]{@{}c@{}}Detection of vehicle\\ traffic existence\end{tabular}                & Coarse                                                             \\ \hline
CCTV2                                                         & \begin{tabular}[c]{@{}c@{}}Snapshots that we extracted from video \\ streams of CCTV cameras monitoring \\ traffic at intersections in the US\end{tabular} & 5K                                                               & High                                                                         & \begin{tabular}[c]{@{}c@{}}Identification of number\\ of vehicles\end{tabular}                  & Fine                                                               \\ \hline
\end{tabular}%
}
\end{minipage}%
\begin{minipage}{.25\linewidth}
\caption{Hashing times for varying numbers of LSH tables.}
\vspace{-0.15cm}
\label{tab:hashing}
\centering
\begin{tabular}{|c|c|}
\hline
\multicolumn{1}{|l|}{\textbf{\begin{tabular}[c]{@{}c@{}}Number of \\ LSH Tables\end{tabular}}} & \textbf{\begin{tabular}[c]{@{}c@{}}Hashing \\ Time (ms)\end{tabular}} \\ \hline
1 LSH Table                                         & 0.4                        \\ \hline
5 LSH Tables                                        & 1.7                        \\ \hline
10 LSH Tables                                       & 3.3                        \\ \hline
\end{tabular}
\end{minipage}
\vspace{-0.6cm}
\end{table*}




\noindent \textbf{\sol prototype implementation:} We developed a \sol prototype 
that includes 1K lines of code (with C++ and Python components) and uses the ndn-cxx library~\cite{ndn-cxx}, which offers software abstractions for NDN communication. 
Our prototype consists of mechanisms running on user devices, the edge network, and \ENs. On devices, we implemented task offloading and LSH mechanisms. Devices select images from the used datasets as task input data, and utilize the FALCONN library to generate the images' locality sensitive hashes. Subsequently, they create tasks and offload them to the edge. In the edge network, we implemented reuse aware mechanisms for task forwarding and the extended Interest pipeline (Figure~\ref{Figure:forwarding}) in the NDN Forwarding Daemon (NFD)~\cite{nfd-dev}, the de facto NDN software forwarder. On \ENs, we implemented mechanisms to execute received tasks, store their execution results, and search for previously executed similar tasks 
based on the search mechanisms of the FALCONN library.

\subsection{Evaluation of \sol Design Components}
\label{subsec:comps}

\noindent \textbf{Setup and metrics:} 
First, we study the extended Interest pipeline implemented in NFD by assessing: (i) the processing time for NFD to forward a task through rFIB compared to forwarding an Interest through FIB as we vary the number of entries and the length of name prefixes per entry for rFIB and FIB respectively; and (ii) the size of rFIB as we increase the number of available \ENs, services, and LSH tables. Second, we study the feasibility of using the FALCONN library to realize LSH in \sol by assessing: (i) the time needed to generate the locality sensitive hash of the task input data; (ii) the time needed to find a previously executed task for the same service with input data that has the highest similarity (among stored previous tasks) to the input data of an incoming task (nearest neighbor search); and (iii) the accuracy of the search process (nearest neighbor search) in the sense of finding a previously executed task for the same service with input data that matches the input data of an incoming task (\eg images of the same object captured from different angles for an object identification service). We perform our evaluation on a low-end desktop computer equipped with an Intel Core i5-4250U CPU@1.30GHz and 8GB of memory to obtain results for the performance of these components on low-end devices. 

\noindent \textbf{Results:} In Figure~\ref{Figure:processing}, we present the processing time for tasks and regular Interests. 
The results demonstrate that the processing of a task through rFIB incurs minimal additional time overhead (up to 5$\mu$s) compared to the processing of an Interest through FIB. This overhead is imposed only once per offloaded task, as we discussed in Section~\ref{subsec:reuse_net}, since a single rFIB lookup is needed per task. The rFIB size remains modest as we increase the number of LSH tables, \ENs, rFIB entries, and available services, as well as the length of name prefixes. Specifically, the rFIB size does not exceed 54.2MB for up to 10 LSH tables, 100 \ENs, 100K rFIB entries, 1K services, 256-byte name prefixes, and an index of 4 bytes per LSH table (the maximum size supported by FALCONN).

In Table~\ref{tab:hashing}, we present the average hashing times through FALCONN for all datasets. The results show that the hashing time increases as the number of the used LSH tables increases. Our results demonstrate that hashing times for images can be practical for up to 10 LSH tables. However, our results in Table~\ref{tab:accuracy} demonstrate that the accuracy of FALCONN shows minor improvements as we increase the number of LSH tables from 5 to 10. Our experiments with images of all datasets (Tables~\ref{tab:hashing},~\ref{tab:accuracy}, and~\ref{tab:search}) consistently show that 1 to 5 LSH tables provide the best tradeoff among hashing times, search times, and accuracy, achieving: (i) satisfactory accuracy with a single LSH table for the MNIST and Stanford AR datasets and with 5 LSH tables for the rest of the datasets; and (ii) hashing times of less than 1.7ms for up to 5 LSH tables.

\begin{table}[t]
\tiny
\caption{LSH nearest neighbor search accuracy (no similarity threshold applied) and search times.}
\vspace{-0.2cm}
\begin{subtable}{.49\columnwidth}
\caption{Search accuracy.}
\vspace{-0.2cm}
\label{tab:accuracy}
\centering
\begin{tabular}{|c|c|c|c|}
\hline
\multicolumn{4}{|c|}{LSH Nearest Neighbor Search Accuracy (\%)}                                                                                                                                                                            \\ \hline
\textbf{Dataset} & \textbf{\begin{tabular}[c]{@{}c@{}}1 LSH \\ Table\end{tabular}} & \textbf{\begin{tabular}[c]{@{}c@{}}5 LSH \\ Tables\end{tabular}} & \textbf{\begin{tabular}[c]{@{}c@{}}10 LSH \\ Tables\end{tabular}} \\ \hline
MNIST            & 83.43                                                           & 84.72                                                            & 86.27                                                             \\ \hline
Pandaset         & 71.54                                                           & 84.48                                                            & 85.21                                                             \\ \hline
Stanford AR      & 97.02                                                           & 98.98                                                            & 98.98                                                             \\ \hline
CCTV1            & 82.70                                                           & 89.20                                                            & 90.03                                                             \\ \hline
CCTV2            & 90.40                                                           & 97.55                                                            & 98.55                                                             \\ \hline
\end{tabular}
\end{subtable}%
\begin{subtable}{.49\columnwidth}
\caption{Search times.}
\vspace{-0.2cm}
\label{tab:search}
\centering
\begin{tabular}{|c|c|c|c|}
\hline
\multicolumn{4}{|c|}{LSH Nearest Neighbor Search Time (ms)}                                                                                                                                                                                                                                   \\ \hline
\textbf{\begin{tabular}[c]{@{}c@{}}Number of \\ Images (x1000)\end{tabular}} & \textbf{\begin{tabular}[c]{@{}c@{}}1 LSH \\ Table\end{tabular}} & \textbf{\begin{tabular}[c]{@{}c@{}}5 LSH \\ Tables\end{tabular}} & \textbf{\begin{tabular}[c]{@{}c@{}}10 LSH \\ Tables\end{tabular}} \\ \hline
20                                                                           & 0.09                                                            & 1.08                                                             & 1.43                                                              \\ \hline
40                                                                           & 0.10                                                            & 1.70                                                             & 2.21                                                              \\ \hline
60                                                                           & 0.11                                                            & 2.62                                                             & 3.05                                                               \\ \hline
80                                                                           & 0.13                                                            & 3.25                                                             & 3.61                                                              \\ \hline
100                                                                          & 0.22                                                            & 3.92                                                             & 4.40                                                              \\ \hline
\end{tabular}
\end{subtable}%
\vspace{-0.3cm}
\end{table}

\subsection {Real-World Experiments with \sol} 
\label{subsec:realworld}

\begin{figure}[!t]
\vspace{-0.15cm}
\begin{minipage}[b]{0.49\columnwidth}
 \centering
 \includegraphics[width=0.9\columnwidth]{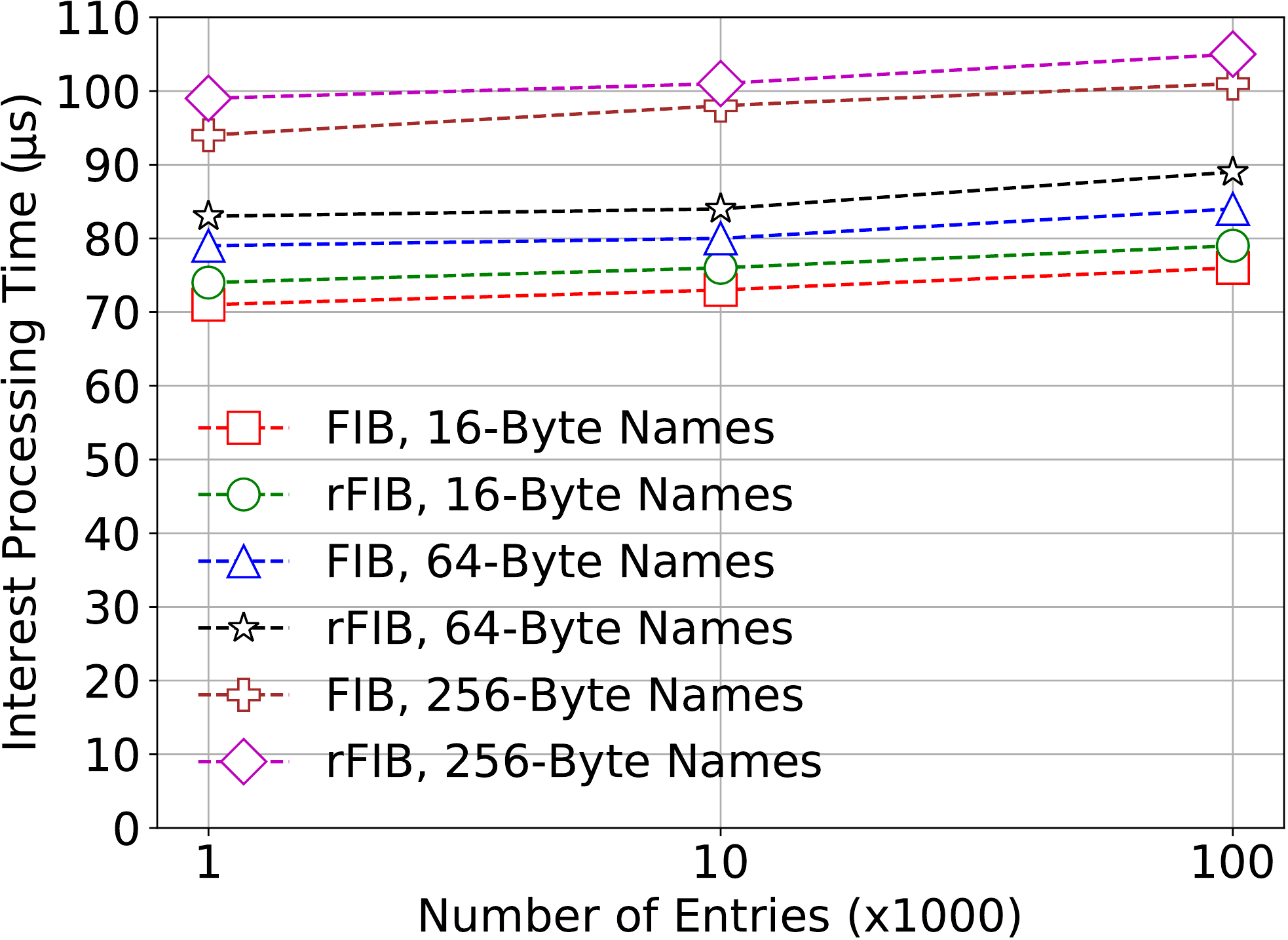}
\vspace{-0.2cm}
 \caption{Processing time for forwarding a task (through rFIB) and an Interest (through FIB).}
 \label{Figure:processing}
 \vspace{-0.4cm}
\end{minipage}
\begin{minipage}[b]{0.485\columnwidth}
 \centering
 \includegraphics[width=1\columnwidth]{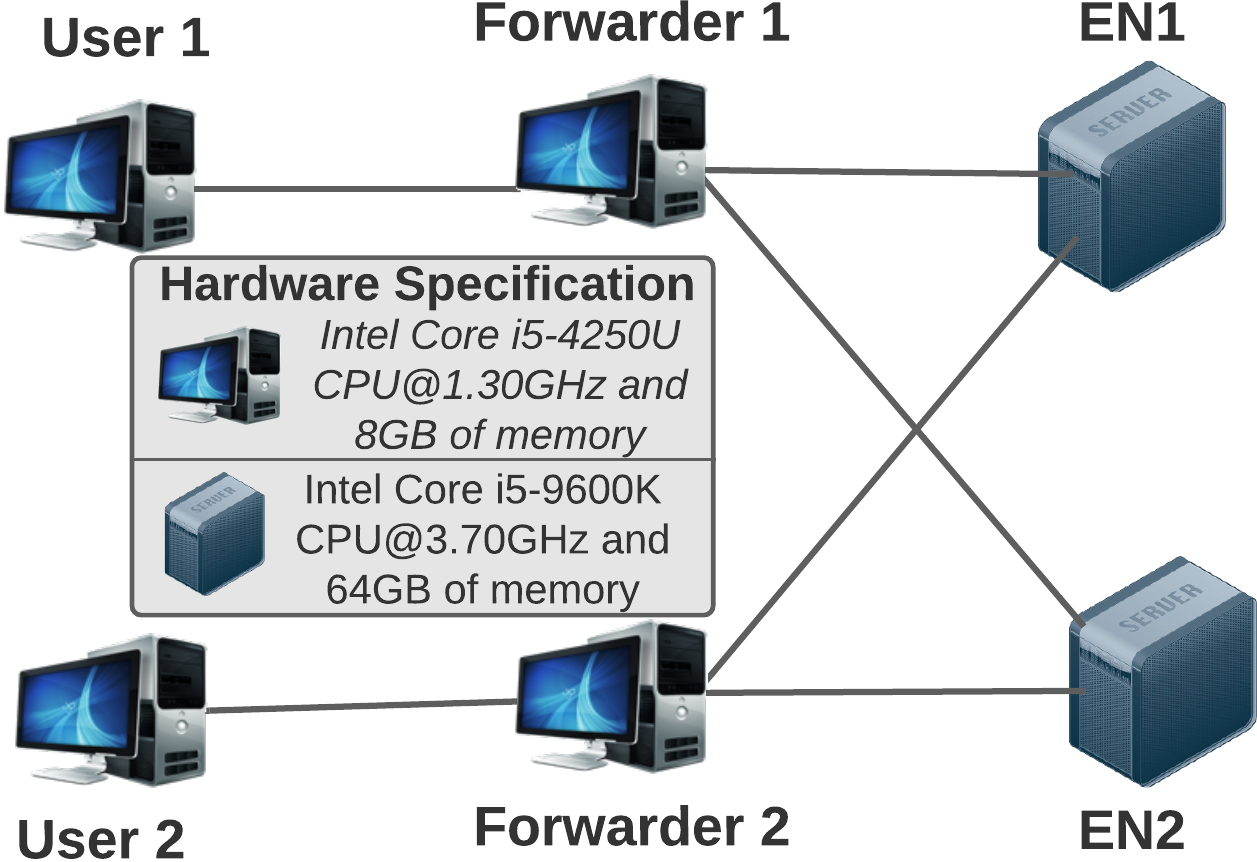}
 \caption{Experimental topology.}
 \label{Figure:topology}
\end{minipage}
\vspace{-0.2cm}
\end{figure}

\noindent \textbf{Setup:} To evaluate the \sol design as a whole, we deploy our prototype on a small-scale testbed (Figure~\ref{Figure:topology}). Our setup consists of two desktop computers acting as user devices that offload tasks to \ENs, two desktop computers acting as NDN forwarders, and two servers acting as \ENs and running tensorflow-based machine learning models~\cite{abadi2016tensorflow}. 
We connect the desktop computers and servers through UDP tunnels and we run NDN as an overlay on top of UDP/IP. The Round Trip Time (RTT) from users to \ENs ranges between 13-21ms with an average value of 18ms. We use a 4-byte index per LSH table and we equally distribute the LSH buckets between the \ENs. We use 1 LSH table for the MNIST and Stanford AR datasets and 5 LSH tables for all other datasets. We measure: (i) the completion time of tasks (\ie the time between offloading a task and receiving its execution results) when the results of previous tasks are reused from the CS of forwarders and the \ENs, and when a task is executed from scratch; (ii) the accuracy of reuse (\ie whether the execution results of a reused task and the results of an offloaded task would have been the same, if the offloaded task had been executed from scratch); and (iii) the percent of reuse (\ie the percent of offloaded tasks that reused the results of previous tasks).

\begin{figure*}[!t]
    \captionsetup[subfigure]{aboveskip=-0.1pt,belowskip=-0.1pt}
	\centering
	\begin{subfigure}{0.24\textwidth}
		\centering
		\includegraphics[width=1\textwidth]{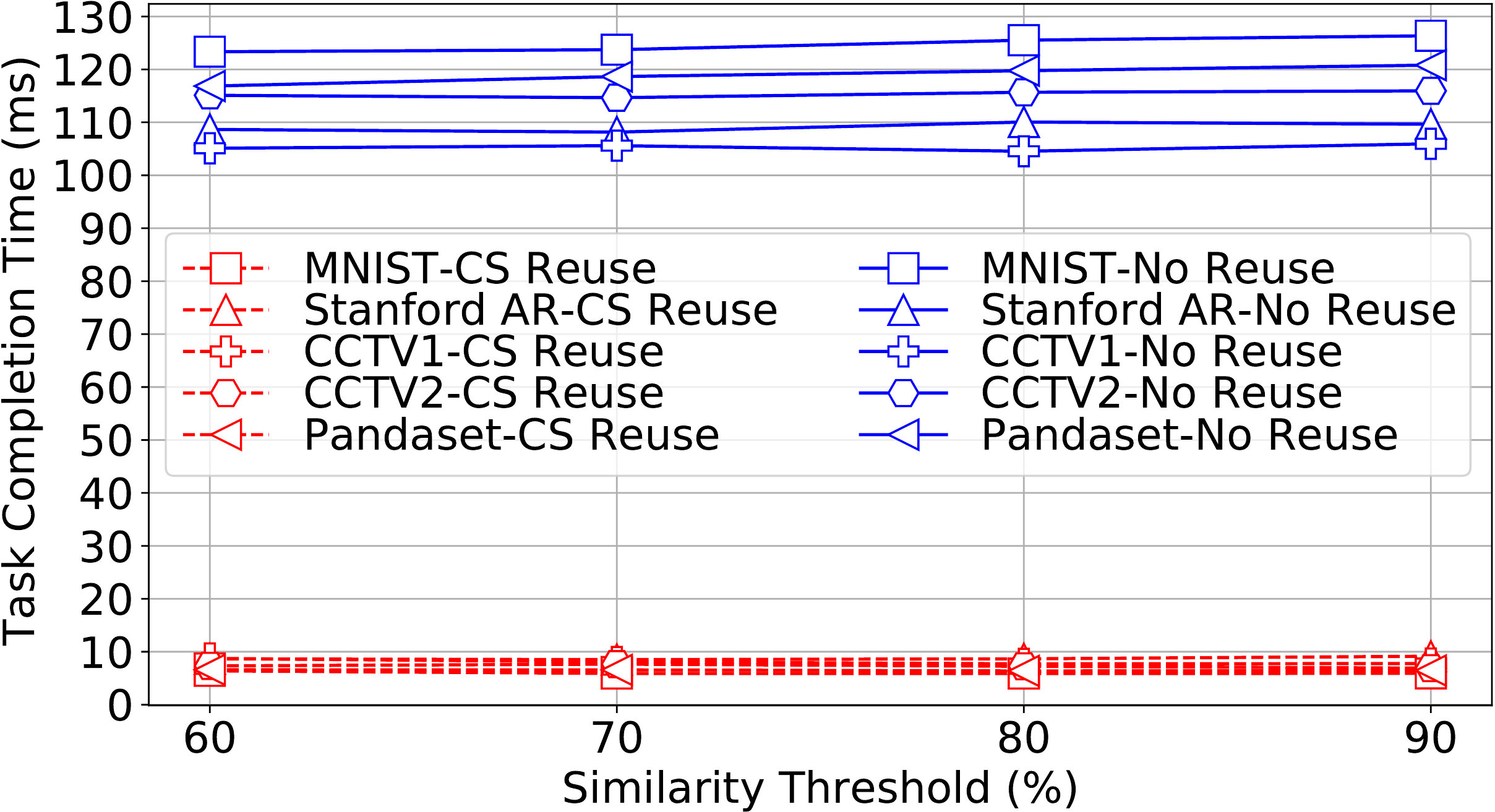}
		\caption{Task completion time with reuse from CS of forwarders and task execution from scratch.} \hfill
		\label{Figure:timecs}
	\end{subfigure} 
	\hfill
	\begin{subfigure}{0.24\textwidth}
		\centering
		\includegraphics[width=1\textwidth]{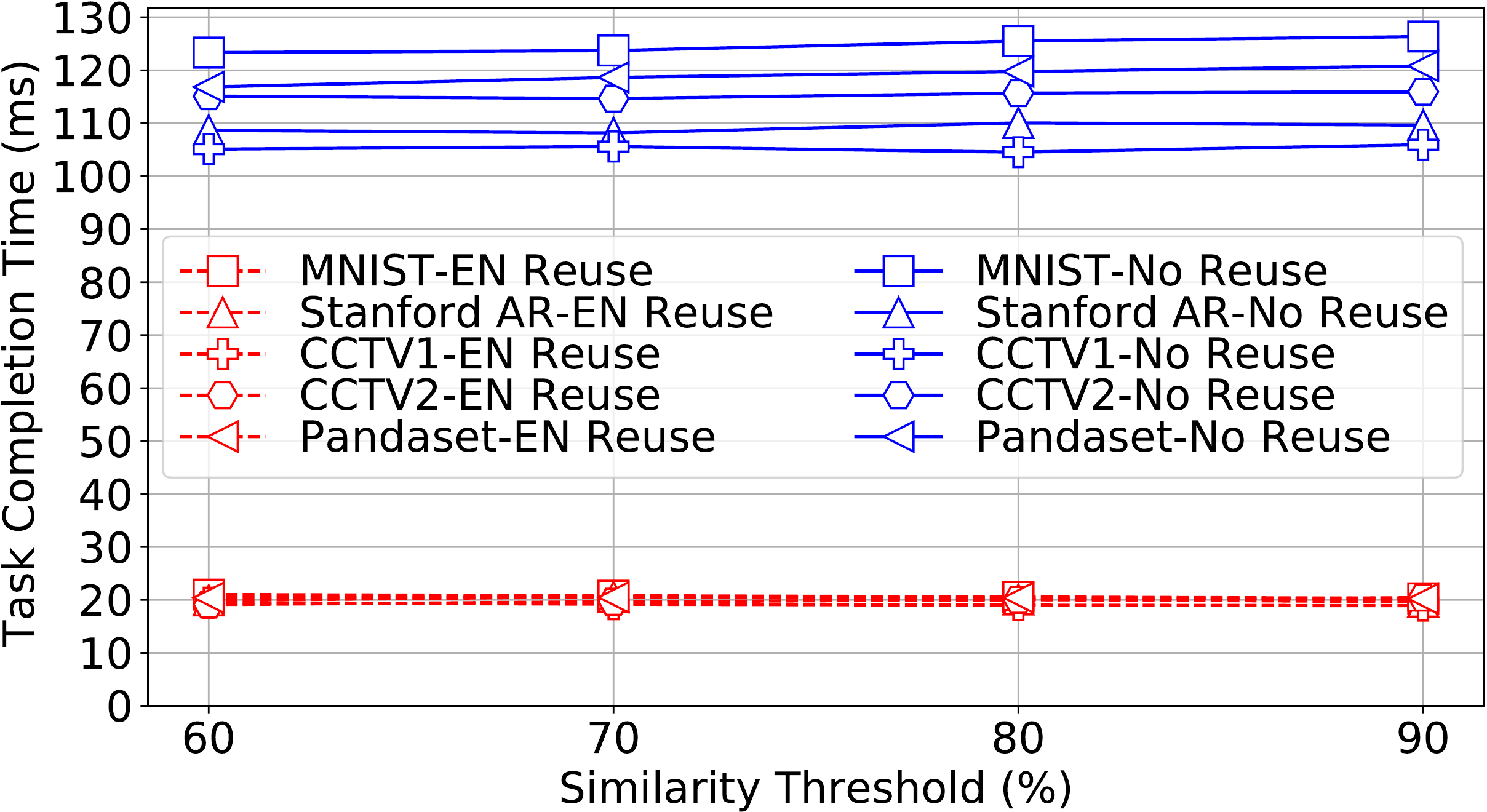}
		\caption{Task completion time with reuse from \ENs and task execution from scratch.}  \hfill
		\label{Figure:timecn}
	\end{subfigure} 
	\hfill
	\begin{subfigure}{0.24\textwidth}
		\centering
		\includegraphics[width=1\textwidth]{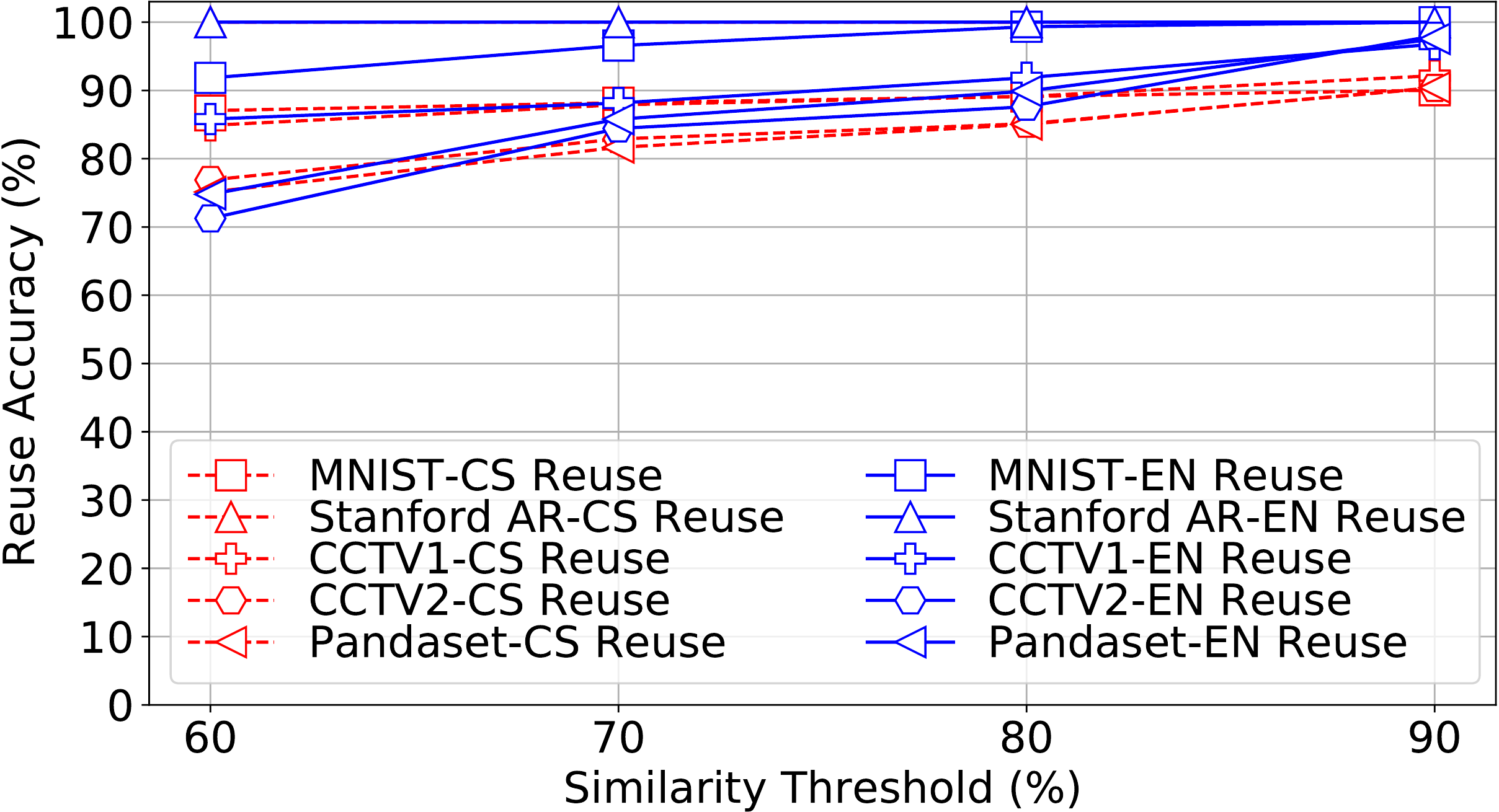}
		\caption{Accuracy of reuse for all datasets and varying similarity thresholds.}  \hfill
		\label{Figure:accuracy}
	\end{subfigure}
	\hfill
	\begin{subfigure}{0.24\textwidth}
		\centering
		\includegraphics[width=1\textwidth]{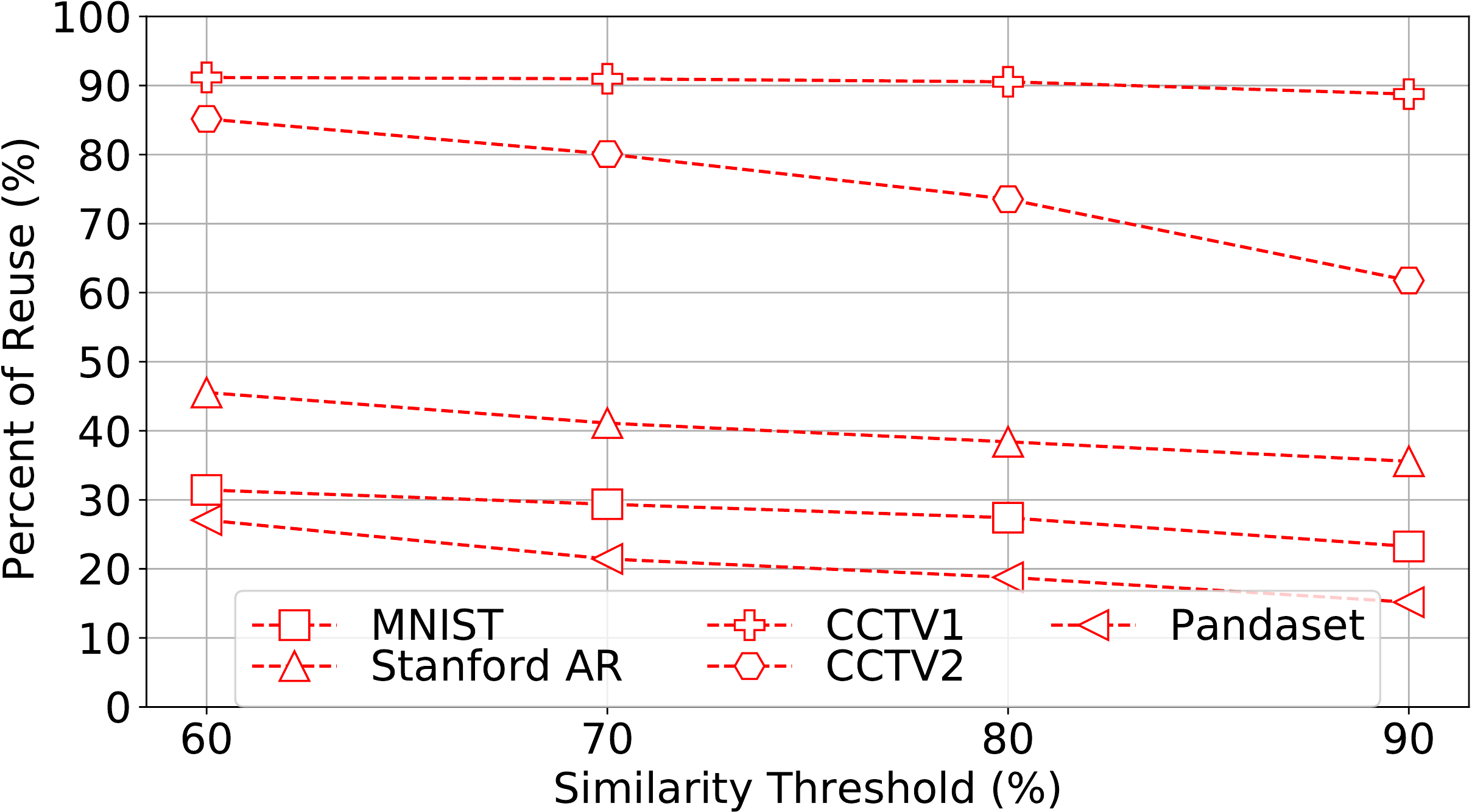}
		\caption{Percent of reuse for for all datasets and varying similarity thresholds.}  \hfill
		\label{Figure:percentreuse}
	\end{subfigure}
	\vspace{-0.45cm}
	\caption{Experimental evaluation results of the \sol prototype.}
	\vspace{-0.5cm}
\end{figure*}

\noindent \textbf{Results:} In Figures~\ref{Figure:timecs} and~\ref{Figure:timecn}, we present the task completion time when tasks are reused from the CS of NDN forwarders and \ENs respectively in comparison with the time for tasks to be executed from scratch for varying similarity thresholds. Our results show that reuse can significantly reduce task completion times. Reusing tasks cached in the CS of forwarders results in 12.02-21.34$\times$ lower task completion times than executing tasks from scratch at \ENs. Furthermore, the reuse of previously executed tasks at \ENs results in 5.25-6.22$\times$ lower task completion times than executing tasks from scratch.

In Figure~\ref{Figure:accuracy}, we present the accuracy of reuse for all datasets and varying similarity thresholds. Depending on the dataset, our results show that \sol can achieve 90-100\% and 96-100\% accuracy for reused tasks cached in the CS of forwarders and stored at the \ENs respectively as we increase the similarity threshold. Our results also indicate that applications requiring fine-grained processing of offloaded data (\eg Pandaset and CCTV2 datasets) may need higher similarity thresholds to achieve reuse accuracy of at least 90\%. On the other hand, applications that require more coarse-grained processing of offloaded data may be able to achieve reuse accuracy of at least 90\% for lower similarity thresholds.


In Figure~\ref{Figure:percentreuse}, we present the percent of reuse for all datasets and varying similarity thresholds. Our results  demonstrate that \sol can reduce the number of tasks that are executed from scratch by an average of 52\% across all the datasets as compared to cases without computation reuse. We further verified that \sol is able to take advantage of virtually all available reuse opportunities for datasets with different degrees of correlation and, therefore, applications offloading tasks with input data that exhibits high, moderate, and low correlation. Especially in cases, where the majority of the input data of tasks is similar (\eg CCTV1 and CCTV2 datasets, which include consecutive snapshots from video streams of vehicle traffic), \sol can reduce the number of tasks that are executed from scratch by up to 91\%. In such cases, up to 27\% and 64\% of the total number of offloaded tasks are satisfied by reusing the execution results of tasks cached in the CS of NDN forwarders and stored at the \ENs respectively. 

Overall, our results signify a trade-off between the percent and the accuracy of reuse; to achieve higher accuracy (Figure~\ref{Figure:accuracy}), applications might need to select higher similarity thresholds, which in turn may reduce the percent of reuse (Figure~\ref{Figure:percentreuse}). Ultimately, the extent of this trade-off depends on the degree of correlation 
between the input data of tasks and the type/granularity of the required input data processing.

\subsection{Network Simulations with \sol} 
\label{subsec:simulations}

\noindent \textbf{Setup:} To scale up our evaluation, we ported our \sol prototype into ndnSIM~\cite{mastorakis2017evolution}, the de facto NDN simulator for ns-3~\cite{henderson2008network}. ndnSIM features integration with the real-world NFD codebase to offer high fidelity of simulation results. We used NetworkX~\cite{hagberg2008exploring} to generate 50 network topologies that resemble autonomous systems on the Internet~\cite{elmokashfi2010scalability}. Each topology consists of 20 to 40 nodes that are interconnected through links with 5ms of delay each, while 10 of these nodes are randomly selected to serve as \ENs. Users are attached to topology nodes through links with 2ms of delay and offload tasks towards the \ENs. We run NDN directly on top of the link layer. Following the setup of our real-world experiments, we use a 4-byte index per LSH table and we equally distribute the LSH buckets among \ENs. We use 1 LSH table for the MNIST and Stanford AR datasets and 5 LSH tables for all other datasets. To make our simulations maximally realistic, we use the following processing delays (measured during the experiments of Sections~\ref{subsec:comps} and~\ref{subsec:realworld}): (i) delays between 71-101$\mu$s and 74-106$\mu$s for an NDN forwarder to process and forward a task through FIB and rFIB respectively; 
(ii) hashing times for 1 and 5 LSH tables as reported in Table~\ref{tab:hashing}; 
(iii) LSH search times for 1 and 5 LSH tables as reported in Table~\ref{tab:search}; 
and (iv) the time for our tensorflow-based machine learning models to process an image ranging between 70-100ms. 
In addition to the metrics we measured in Section~\ref{subsec:realworld}, we define and measure the \emph{task forwarding error rate} as the percent of tasks that are forwarded to an \EN that does not have a similar task to reuse, however, such a similar task is stored at another \EN. This can occur when multiple LSH tables are maintained for a service, since the network forwards tasks towards the \EN that handles the majority of the indexed buckets to maximize the chances of reuse as described in Section~\ref{subsec:reuse_net}.

\begin{figure*}[t]
    \captionsetup[subfigure]{aboveskip=-0.1pt,belowskip=-0.1pt}
	\centering
	\begin{subfigure}{0.24\textwidth}
		\centering
		\includegraphics[width=1\textwidth]{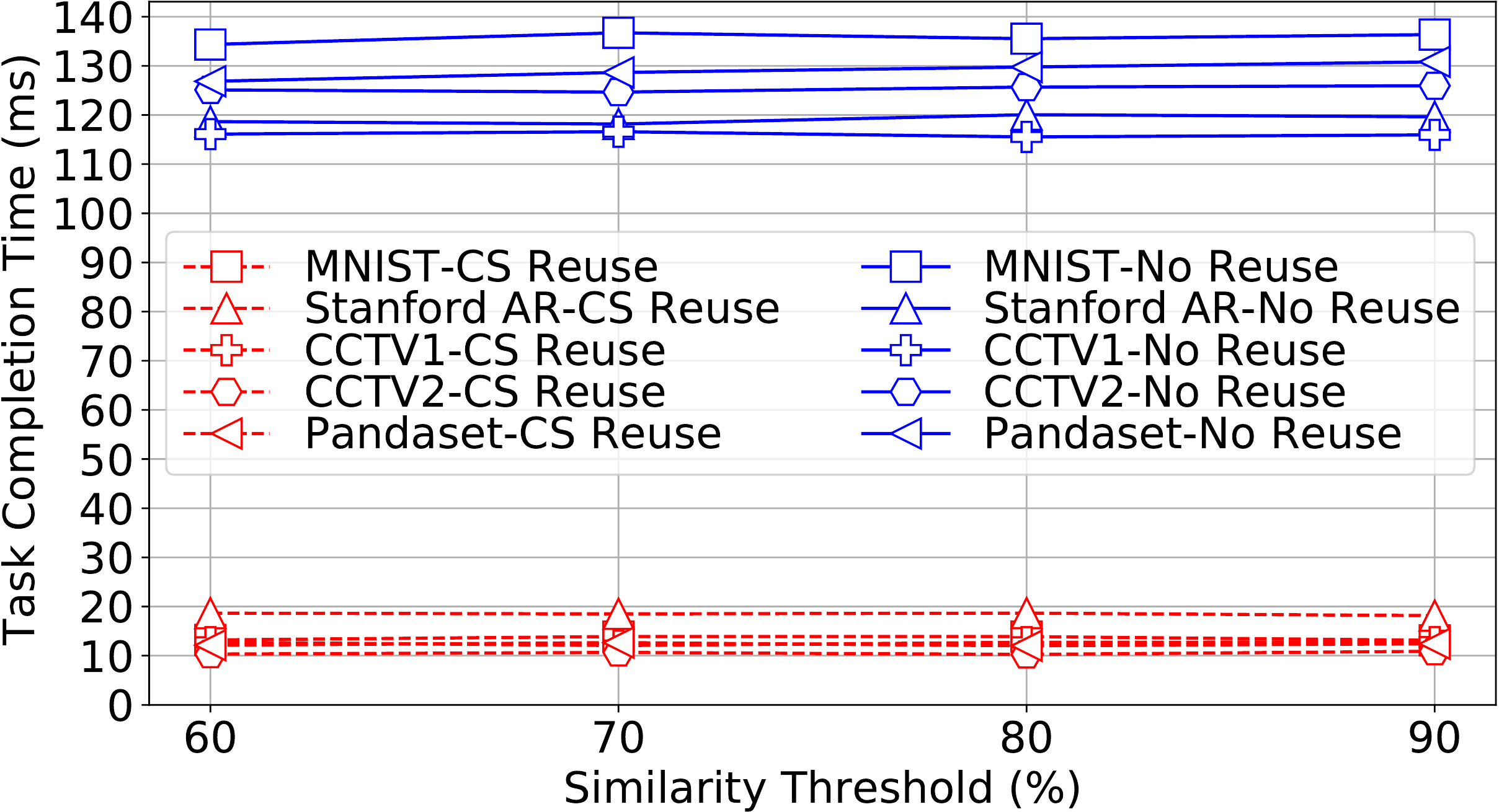}
		\caption{Task completion time with reuse from CS of forwarders and task execution from scratch.} \hfill
		\label{Figure:timecs-sim}
	\end{subfigure} 
	\hfill
	\begin{subfigure}{0.24\textwidth}
		\centering
		\includegraphics[width=1\textwidth]{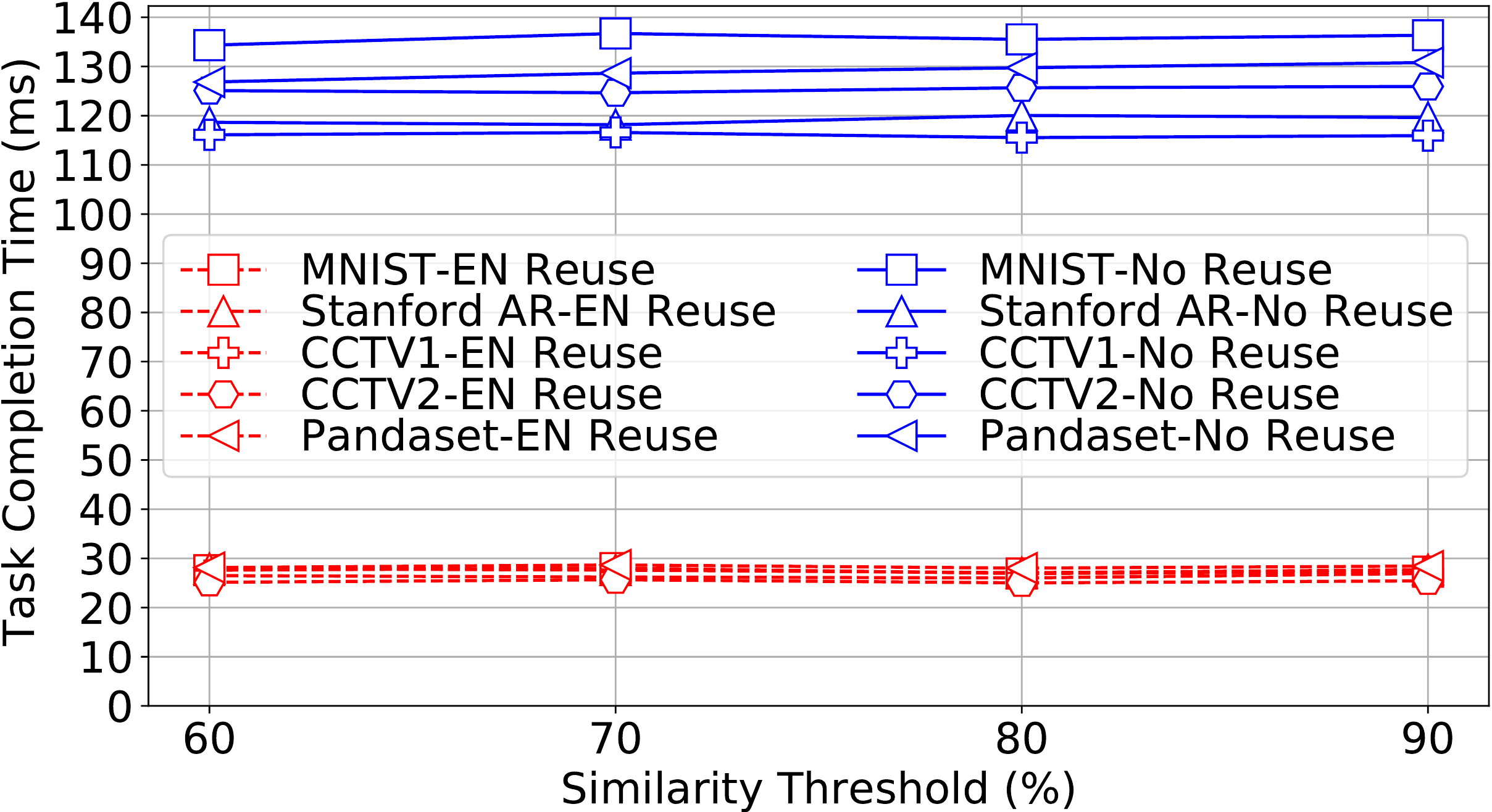}
		\caption{Task completion time with reuse from \ENs and task execution from scratch.} \hfill
		\label{Figure:timecn-sim}
	\end{subfigure} 
	\hfill
	\begin{subfigure}{0.24\textwidth}
		\centering
		\includegraphics[width=1\textwidth]{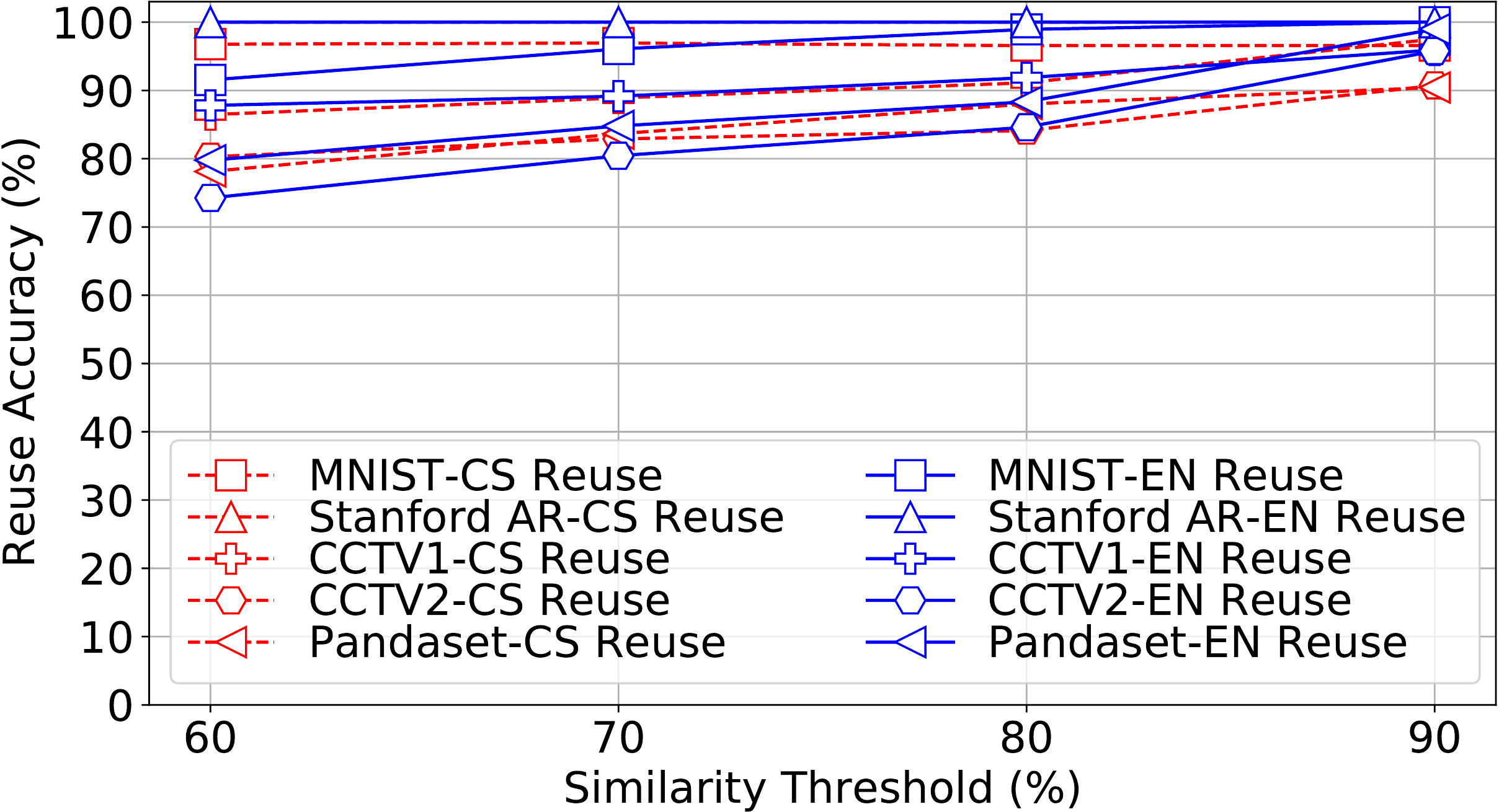}
		\caption{Accuracy of reuse for all datasets and varying similarity thresholds.} \hfill
		\label{Figure:accuracy-sim}
	\end{subfigure}
	\hfill
	\begin{subfigure}{0.24\textwidth}
		\centering
		\includegraphics[width=1\textwidth]{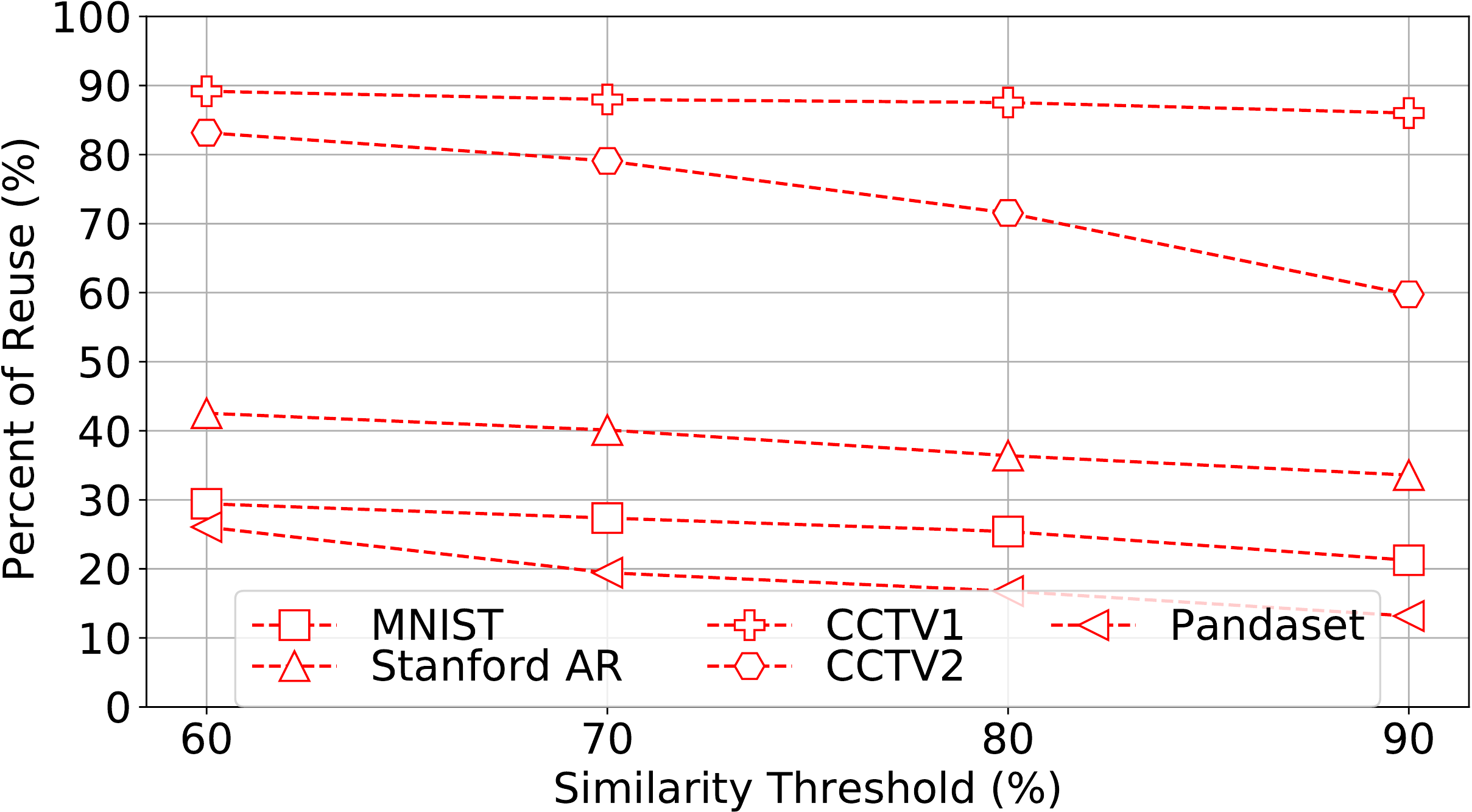}
		\caption{Percent of reuse for all datasets and varying similarity thresholds.} \hfill
		\label{Figure:percentreuse-sim}
	\end{subfigure}
	\vspace{-0.5cm}
	\caption{Simulation evaluation results of \sol.}
	\vspace{-0.6cm}
\end{figure*}

\noindent \textbf{Results:} In Figures~\ref{Figure:timecs-sim} and~\ref{Figure:timecn-sim}, we present the task completion time when tasks are reused from the CS of NDN forwarders and \ENs respectively in comparison to the time needed for tasks to be executed from scratch by \ENs. Our results demonstrate the same trend as the experimental results of Section~\ref{subsec:realworld}: \sol achieves 6.43-12.28$\times$ and 4.25-5.11$\times$ lower task completion times when the results of previously executed tasks are reused from the CS of forwarders and \ENs respectively in comparison to executing tasks from scratch. The reduction magnitude of the task completion times due to reuse is lower in our simulation results than the experimental results of Section~\ref{subsec:realworld}. This is due to having \ENs up to 4 hops away from users, thus the network paths become longer and the network delay may increase for the completion of tasks. 

In Figure~\ref{Figure:accuracy-sim}, we present results on the accuracy of reuse, which demonstrate the same trend as the experimental results of Section~\ref{subsec:realworld}. \sol can achieve 90-100\% and 96-100\% accuracy for reused tasks cached in the CS of forwarders and stored at the \ENs respectively as we increase the similarity threshold. In Figure~\ref{Figure:percentreuse-sim}, we present results on the percent of reuse, which demonstrate the same trend as the results of Section~\ref{subsec:realworld}. \sol can reduce the number of tasks executed from scratch by an average of 50\% among all datasets as compared to cases without computation reuse. When the majority of task input data is similar (CCTV1 and CCTV2 datasets), \sol can reduce the number of tasks executed from scratch by up to 88\%. In such cases, up to 25\% and 63\% of the total number of tasks reuse the results of tasks cached in the CS of forwarders and stored at the \ENs respectively. 

We further conducted experiments for variable cache/storage sizes at user devices, NDN forwarders (edge network), and \ENs considering a Least Recently Used (LRU) cache replacement policy. Our results indicate that the percent of reuse increases until we reach cache sizes that can hold all the offloaded tasks that 
need to be executed from scratch. Increasing the size of caches beyond this point has no impact on the percent of reuse. Our results also indicate that the impact of cache sizes on the percent of reuse depends both on the degree of similarity between task input data and the minimum similarity threshold, however, the degree of similarity (correlation) between input data has the strongest impact. In terms of reuse accuracy, as we increase the cache size and until we reach the point where all tasks that need to be executed from scratch can fit into caches, the accuracy of reuse decreases. This is due to reusing a larger volume of computation, thus \sol could select to reuse previous tasks, which would not yield the same results as the corresponding offloaded tasks.

\begin{figure}[!tbp]
  \centering
  \captionsetup[subfigure]{aboveskip=-0.00000000001pt,belowskip=-0.00000000001pt}
  \vspace{-0.1cm}
  \begin{minipage}[b]{0.49\columnwidth}
    \includegraphics[width=1\columnwidth]{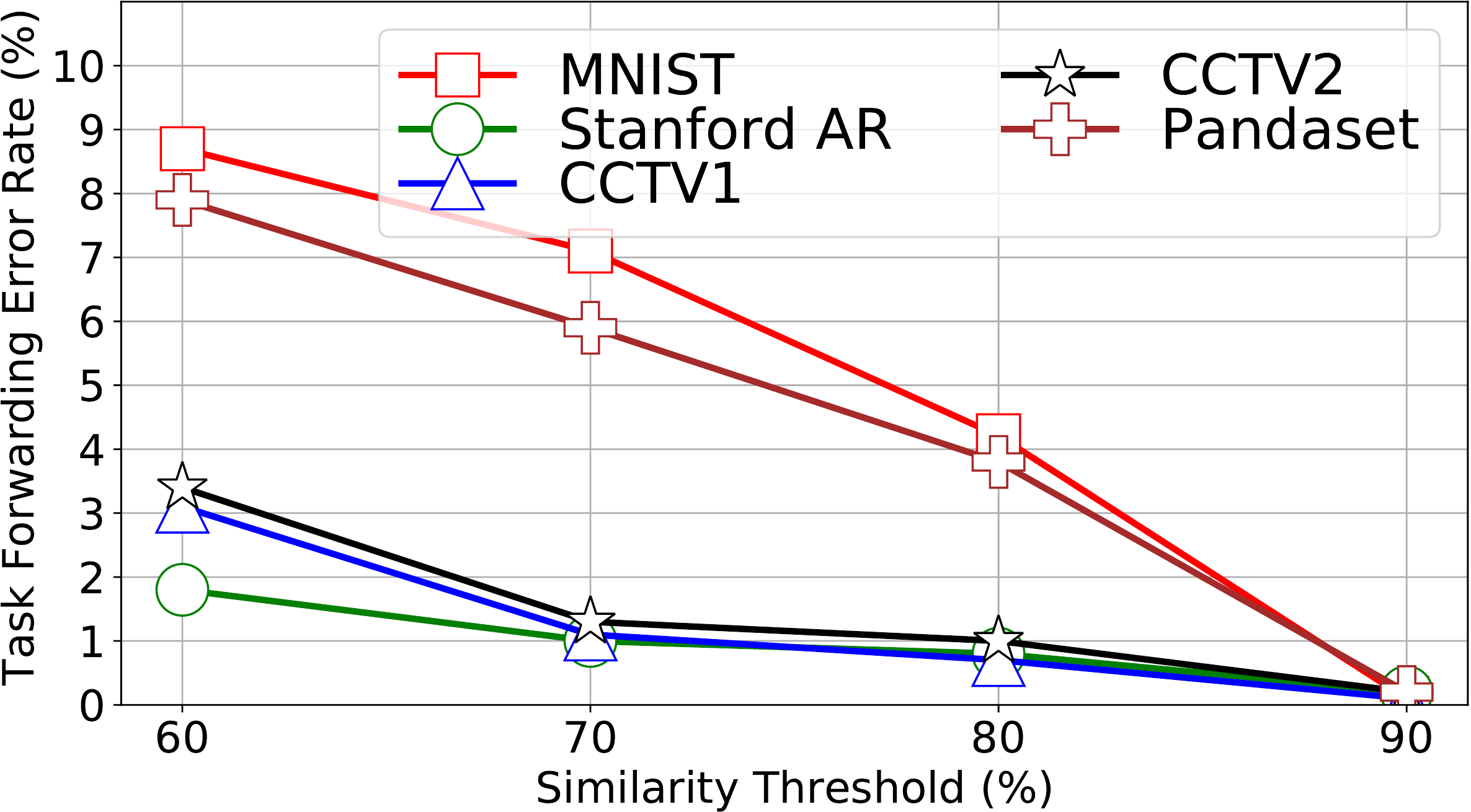}
    \caption{Task forwarding error rate results.} 
    \label{Figure:errorrate}
  \end{minipage}
  \hfill
  \begin{minipage}[b]{0.49\columnwidth}
    \includegraphics[width=1\columnwidth]{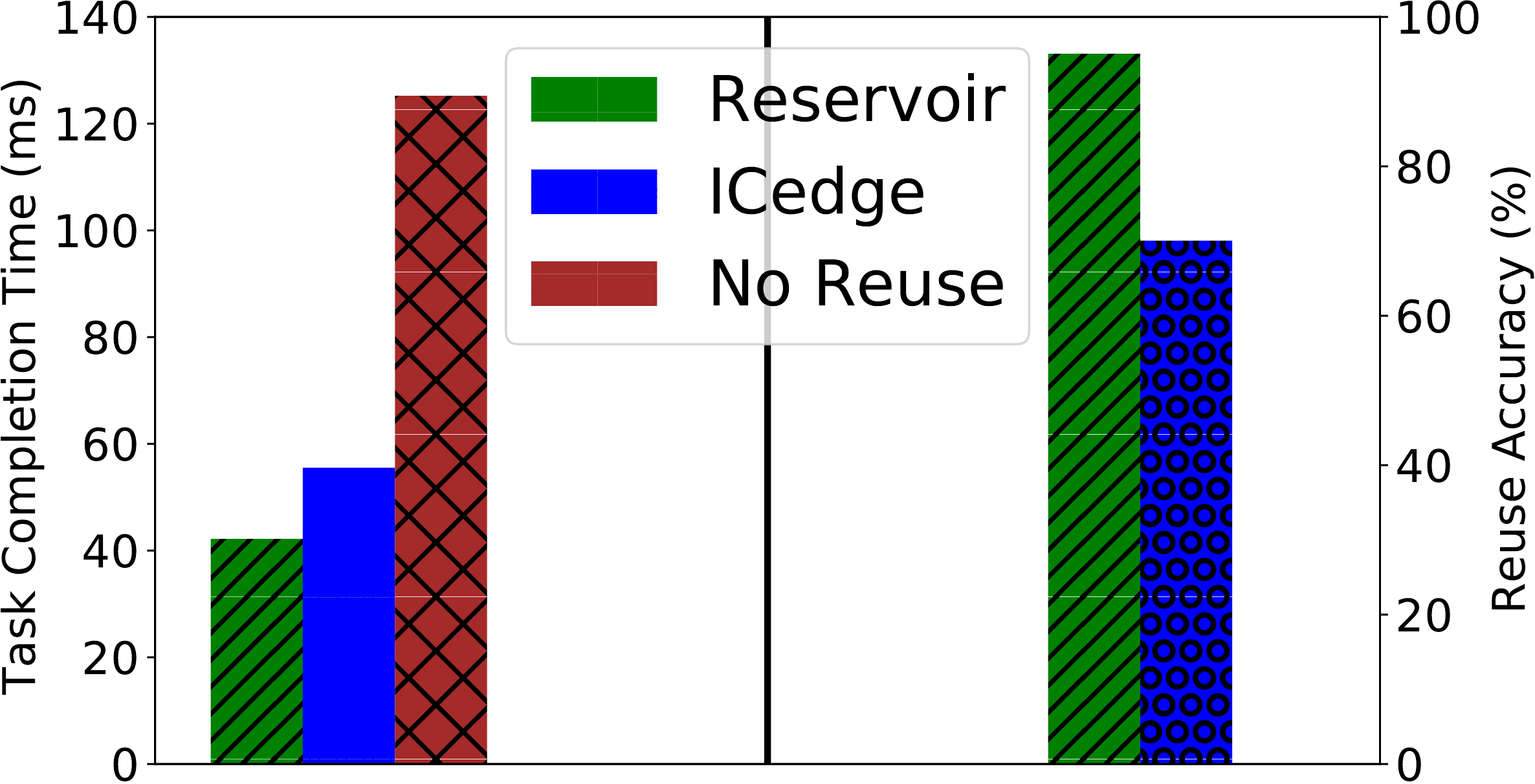}
    \vspace{-0.2cm}
    \caption{Comparison between \sol and ICedge.}
    \label{Figure:comparison}
  \end{minipage}
  \vspace{-1cm}
\end{figure}

In Figure~\ref{Figure:errorrate}, we present the task forwarding error rate results. 
To obtain these results, we used 5 LSH tables for each dataset. Our results show that the task forwarding error rates are lower than 9\% for all datasets and similarity thresholds. As we increase the similarity threshold, the error rates decrease, since the number of previous tasks with input data similar to the input data of incoming tasks decreases, thus becoming less likely that similar tasks may be stored across different \ENs.

\subsection{Comparison to ICedge}

\noindent \textbf{Setup:} We compared the processing time for forwarding a task through \sol and ICedge~\cite{mastorakis2020icedge}. We also conducted network simulations with \sol and ICedge based on the topology and parameters of Section~\ref{subsec:simulations} to quantify and compare the completion time of tasks and the reuse accuracy.

\noindent \textbf{Results:} ICedge requires 77-111$\mu$s of processing time to forward a task, since it employs both name-based lookups through FIB and a specialized forwarding mechanism per application. This processing time is 6-10$\mu$s higher than the processing time required by \sol. In Figure~\ref{Figure:comparison}, we present the average task completion time and reuse accuracy among all datasets for \sol and ICedge. \sol achieves about 24\% lower task completion times than ICedge, since it takes full advantage of in-network caching to retrieve the results of similar tasks from the CS of forwarders whenever possible. \sol also achieves 26\% higher reuse accuracy than ICedge. ICedge identifies reuse opportunities through naming semantics. Such semantics provide limited information about the task input data (\eg the location where a picture was taken, the device that took a picture), without providing any indication of the similarity among tasks. On the other hand, the combination of naming with LSH in \sol provides a light-weight mechanism for the accurate identification of similar tasks.




\section{Conclusion}
\label{sec:concl}

In this paper, we presented \sol, a framework to enable pervasive computation reuse in practical edge computing environments. \sol makes the edge network infrastructure aware of the semantics of computation reuse, thus being able to identify and forward similar tasks towards the same \ENs 
in order to minimize the execution of redundant computation. Our evaluation results demonstrated that \sol can effectively reuse computation at the edge of the network, while incurring marginal performance overheads, achieving up to perfect accuracy, and accommodating applications and computational tasks with different characteristics


\section*{Acknowledgements}

This work is partially supported by National Science Foundation awards CNS-2104700, CNS-2016714, and CBET-2124918, the National Institutes of Health (NIGMS/P20GM109090), the Nebraska University Collaboration Initiative, and the Nebraska Tobacco Settlement Biomedical Research Development Funds.

\bibliographystyle{IEEEtran}
\bibliography{sections/refs.bib}
\end{document}